\documentclass[aps,prb,twocolumn,superscriptaddress,floatfix]{revtex4-1}

\usepackage{float}

\usepackage[utf8]{inputenc}
\usepackage{graphicx}
\usepackage{lmodern}
\usepackage[fleqn]{amsmath} 
\usepackage{amsthm,amsfonts,amssymb}
\usepackage{wasysym}
\usepackage{bbm}            
\usepackage{soul}
\usepackage{xspace}
\usepackage{bm}
\usepackage{dsfont}
\usepackage{footnote}
\usepackage{alltt}
\usepackage{xcolor}
\usepackage{color}
\usepackage{tcolorbox}
\usepackage{listings}
\usepackage{url}
\usepackage{booktabs}
\usepackage[vcentermath]{youngtab}
\usepackage[colorlinks = true,
            linkcolor = blue,
            urlcolor  = blue,
            citecolor = blue,
            anchorcolor = blue]{hyperref}
\usepackage[caption=false]{subfig}

\newcommand*{\bb}{\tilde{b}} 
\newcommand*{\ba}{\tilde{a}} 
\newcommand{\ket}[1]{\left| #1 \right>} 
\definecolor{FKnotecolor}{rgb}{0.1,0.5,0.8}

\definecolor{KPnotecolor}{rgb}{0.9,0.1,0.1}

\begin{document}

\title{Dimensional crossover in the SU(4) Heisenberg model in the six-dimensional antisymmetric self-conjugate representation revealed by quantum Monte Carlo and linear flavor-wave theory}

\author{Francisco H. Kim}
\affiliation{Institute of Physics, \'Ecole Polytechnique F\'ed\'erale de Lausanne (EPFL), CH-1015 Lausanne, Switzerland}
\author{Fakher F. Assaad}
\affiliation{Institut für Theoretische Physik und Astrophysik, Universität Würzburg, Am Hubland, D-97074 Würzburg, Germany}
\affiliation{W\"urzburg-Dresden Cluster of Excellence ct.qmat, Universität Würzburg, Am Hubland, D-97074 Würzburg, Germany } 
\author{Karlo Penc}
\affiliation{Institute for Solid State Physics and Optics, Wigner Research Centre for Physics, Hungarian Academy of Sciences, H-1525 Budapest, P.O.B. 49, Hungary}
\author{Fr\'ed\'eric Mila}
\affiliation{Institute of Physics, \'Ecole Polytechnique F\'ed\'erale de Lausanne (EPFL), CH-1015 Lausanne, Switzerland}

\date{\today}

\begin{abstract}
Using linear flavor-wave theory (LFWT) and auxiliary field quantum Monte Carlo (QMC), we investigate the properties of the SU(4) Heisenberg model on the anisotropic square lattice in the fully antisymmetric six-dimensional irreducible representation, a model that describes interacting fermions with four flavors at half-filling. Thanks to the calculations on very large systems, we have been able to convincingly demonstrate that QMC results are consistent with a small but finite antiferromagnetic moment at the isotropic point, in qualitative agreement with LFWT obtained earlier [F. H. Kim {\it{et al.}}, Phys. Rev. B {\bf 96}, 205142 (2017)], and in quantitative agreement with results obtained previously on the Hubbard model [D. Wang {\it{et al.}}, Phys. Rev. Lett. {\bf 112}, 156403 (2014)] after extrapolation to infinite $U/t$. The presence of a long-range antiferromagnetic order has been further confirmed by showing that a phase transition takes place into a valence-bond solid (VBS) phase not too far from the isotropic point when reducing the coupling constant along one direction on the way to decoupled chains.
\end{abstract}


\maketitle

\section{\label{sec:intro}Introduction}
Substantial leaps of progress have been reported in recent years in the field of ultracold atom manipulation in optical lattices \cite{wu_exact_2003,honerkamp_ultracold_2004,jaksch_cold_2005,cazalilla_ultracold_2009,gorshkov_two-orbital_2010,taie_su6_2012,pagano_one-dimensional_2014,scazza_observation_2014,zhang_spectroscopic_2014,hofrichter_direct_2016}. This, in turn, raises an exciting prospect of realizing the $\mathrm{SU}(N)$ symmetric fermionic Hubbard model with $N$ flavors experimentally, whose Hamiltonian is given by
\begin{equation}
  \label{eq:intro-hubbard}
  \mathcal{H} = -t \sum\limits_{\left\langle i,j \right\rangle,\mu} \left( f ^{\dagger}_{i,\mu} f ^{}_{j,\mu} + \text{H.c.}\right) + U
  \sum\limits_{i,\mu < \nu} n_{i,\mu} n_{i,\nu}
\end{equation}
with fermionic operators $f ^{\dagger}_{i,\mu}$, $f ^{}_{i,\mu}$ on each site $i$ with $N$ flavors, $n_{i,\mu} = f^{\dagger}_{i,\mu}f^{}_{i,\mu}$, and the summation is over the $\left\langle i,j \right\rangle$ nearest-neighbor sites. In the Mott insulating phase, the low-energy physics of this model in the second order in $t/U$ is captured by the antiferromagnetic $\mathrm{SU}(N)$ Heisenberg model,
\begin{equation}
  \label{eq:intro-H}
  \mathcal{H} = J \sum\limits_{\left\langle i,j \right\rangle} \sum\limits_{\mu,\nu} \hat{S} ^{\mu}_{\nu}(i) \hat{S}
  ^{\nu}_{\mu}(j)
\end{equation}
where the operators $\hat{S}^{\mu}_{\nu}$ simply exchanges the $\mathrm{SU}(N)$ flavor $\mu$ with $\nu$, and $J\sim t^2/U$. When having one particle per site, the flavor states are described by the fundamental irreducible representation (irrep) of $\mathrm{SU}(N)$ denoted by the Young tableau with one box $\tiny{\yng(1)}$. In contrast, when multiple particles are present per site, the flavor states are described by a different irrep of $\mathrm{SU}(N)$ depending on the flavor symmetry that the particles form. The $\mathrm{SU}(N)$ antiferromagnetic Heisenberg model has been of considerable interest for a while because of the abundance of interesting phases and physical phenomena that it can accommodate. They should soon be within reach, allowing perhaps the realization of a variety of exotic phases shown by recent theoretical and numerical activities in the field \cite{assaad_phase_2005,hermele_mott_2009,corboz_simultaneous_2011,hermele_topological_2011,corboz_spin-orbital_2012,capponi_phases_2016,dufour_stabilization_2016,nataf_plaquette_2016,nataf_chiral_2016,lecheminant_lattice_2017,fuji_non-abelian_2017} thanks to the unprecedented control over various parameters that the optical lattices offer. Notably, one-dimensional systems in the fundamental irrep of $\mathrm{SU}(N)$ were already solved exactly by Sutherland in the 70s using the Bethe ansatz \cite{sutherland_model_1975}, and calculations using the mean-field saddle-point treatment in the large-$N$ limit have been performed for various irreps \cite{affleck_large-n_1988,marston_large-$n$_1989,Read89,read_valence-bond_1989,Read90,read_large-_1991} in the late 80s and early 90s, shedding light on the theoretical understanding of the $\mathrm{SU}(N)$ models in a controlled way. However, the nature of the large-$N$ expansion implies that the validity of its results could be questionable for small values of $N$. Since the enhanced $\mathrm{SU}(N)$ symmetry seems to be physically realizable for up to $N=10$ with up to two particles per site \cite{gorshkov_two-orbital_2010,scazza_observation_2014,zhang_spectroscopic_2014}, it is crucial to have a reliable assessment of these systems with a relatively low $N$.

The model of interest in the present article is the $\mathrm{SU}(4)$ AFM Heisenberg model at half filling (with two fermionic particles per site) in the fully antisymmetric configuration. The states thus belong to the fully antisymmetric self-conjugate representation and this irrep corresponds to the Young tableau $\tiny{\yng(1,1)}$ (two boxes placed in one column). For this model, the large-$N$ limit calculations at zero temperature have long predicted a degenerate dimerized ground state in one dimension \cite{marston_large-$n$_1989,Read89}, with other analytical approaches and numerical methods such as the density matrix renormalization group (DMRG), the quantum Monte Carlo (QMC) and variational Monte Carlo (VMC) calculations also reaching the same conclusion \cite{onufriev_enlarged_1999,assaraf_dynamical_2004,paramekanti_su_2007,nonne_competing_2011}. The Coleman-Mermin-Wagner theorem \cite{mermin_absence_1966,coleman_there_1973} indeed excludes the possibility of having a long-range order in 1D, but this is not the case in 2D at zero temperature. In the two-dimensional square lattice, the Néel-ordered configuration has been suggested as a possible ground state by VMC calculations \cite{paramekanti_su_2007}, and this possibility has been further supported by the linear flavor-wave theory (LFWT) \cite{kim_linear_2017}, an extension of the spin-wave theory for $\mathrm{SU}(2)$ spins. Furthermore, QMC simulations carried out on the $\mathrm{SU}(4)$ Hubbard model in the strong-coupling regime with sizes up to $16 \times 16$ show the Néel ordering \cite{cai_quantum_2013,Wang14b}. However, it remains to be seen if this magnetic order will survive in the Heisenberg limit. As a matter of fact, auxiliary field QMC simulations with system sizes up to $24 \times 24$ seem to suggest the absence of long-range order \cite{assaad_phase_2005} for the $\mathrm{SU}(4)$ AFM Heisenberg model at half filling. The existence of an ordered magnetic state in this model thus appeals for further investigations.

\begin{figure}[t]
  \subfloat[Néel]{\includegraphics[height=3.35cm]{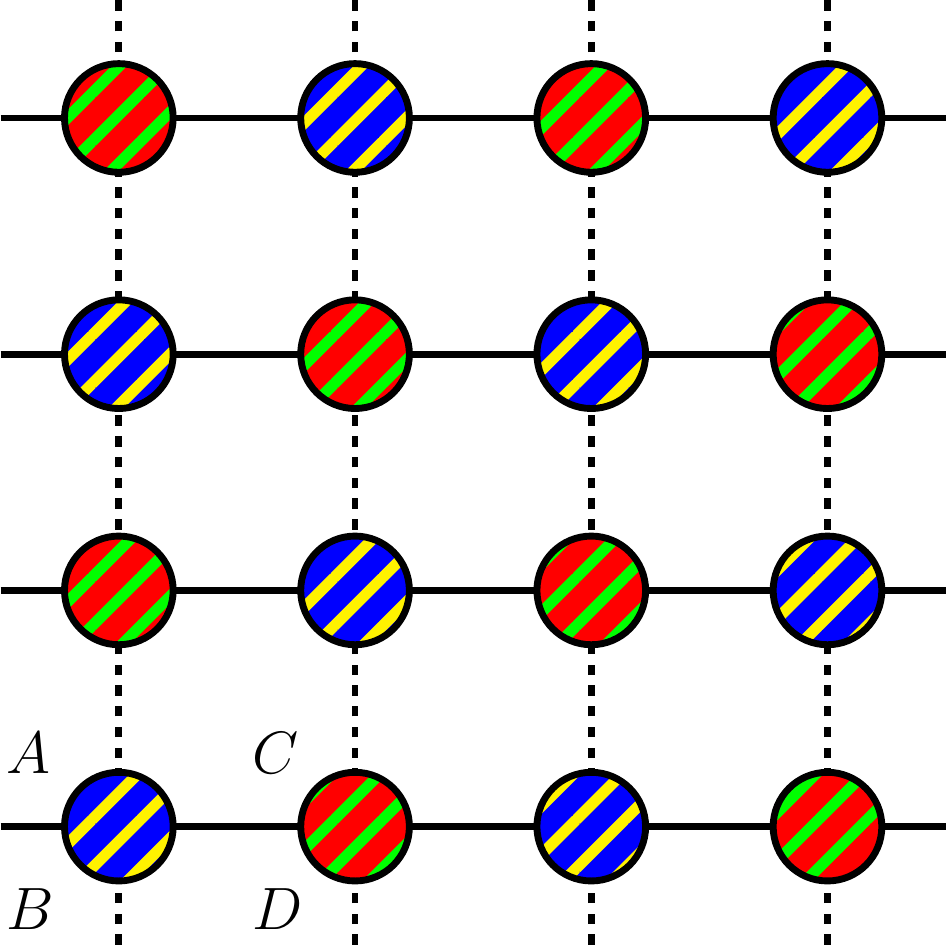}}\hfill
  \subfloat[VBS]{\includegraphics[height=3.35cm]{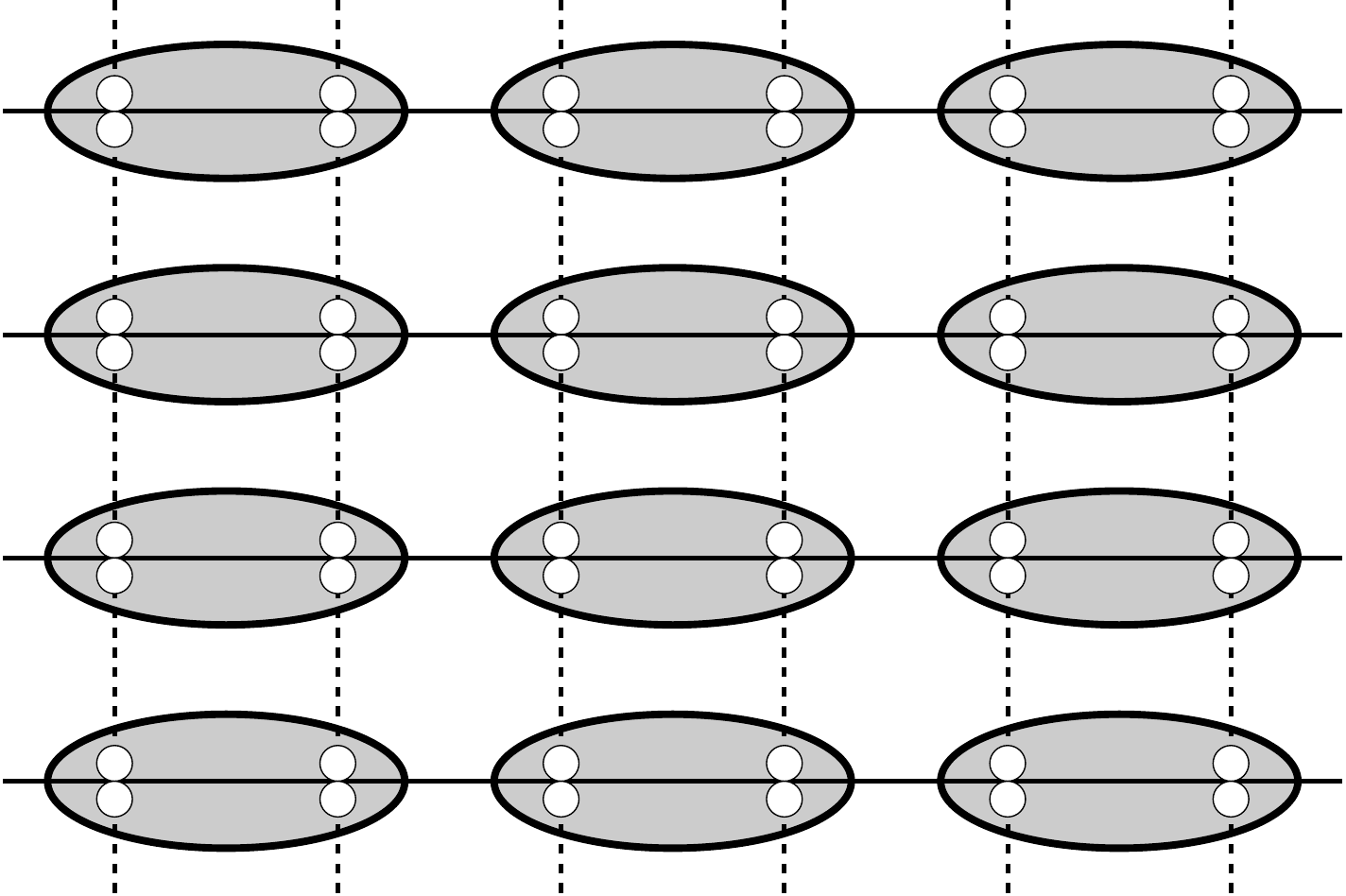}}
  \caption{\label{fig:lattice}Illustrations of the (a) N\'eel-like pattern with an ordering vector $\mathbf{q} = (\pi, \pi)$ and (b) VBS configuration with an ordering vector $\mathbf{q} = (\pi, 0)$ for the Mott-insulating state of SU(4) fermions with two particles per site. The flavors $A,B,C,D$ of the fermions are represented by the colors blue, yellow, red, and green, respectively. The horizontal lines represent the intra-chain coupling $J_{x}$ whereas the vertical dashed lines represent the inter-chain coupling $J_{y}$ that controls the dimensional crossover. The grey ellipse-shape objects in (b) indicate the strongly entangled pairs of sites.}
\end{figure}

To progress further on this issue, we study here the evolution of this system between 2D and 1D by tuning the inter-chain couplings (thus obtaining a collection of 1D chains from the 2D square lattice). The aim is to show that this dimensional crossover triggers a continuous phase transition to a valence bond solid (VBS) in 1D, and that it supports the long-range antiferromagnetic configuration for the 2D lattice, albeit with a small magnetic moment. An example of the N\'eel-like configuration and the VBS configuration is shown in Fig.~\ref{fig:lattice}.
The phase transition from the Néel state during this dimensional crossover will first be assessed by the LFWT by closely following the steps in Ref.~[\onlinecite{kim_linear_2017}]. The results of the auxiliary field QMC simulations (free of the sign problem for the current model) will then be presented by considering system sizes up to $40 \times 40$, showing a small local moment in the 2D model and supporting a continuous transition between the Néel state and the VBS state during the dimensional transition.

\section{\label{sec:lfwt-hamiltonian} The magnetic transition with the Linear Flavor-Wave Theory (LFWT)}
We first define the $\mathrm{SU}(4)$ AFM Heisenberg model in 2D with the intra-chain coupling $J_{x}$ and the inter-chain coupling $J_{y}$ depicted in Fig.~\ref{fig:lattice},
\begin{equation}
  \begin{aligned}
    \label{eq:lfwt-H}
    \mathcal{H} =& \sum\limits_{\langle \vec{\imath}, \vec{\jmath}\rangle} \sum\limits_{\mu,\nu}
    J_{\vec{\imath},\vec{\jmath}} \, \hat{S} ^{\mu}_{\nu}(\vec{\imath}) \hat{S}
    ^{\nu}_{\mu}(\vec{\jmath}).
  \end{aligned}
\end{equation}
The site indices $\langle \vec{\imath}, \vec{\jmath} \rangle$ run over the nearest neighbors, and the indices ${\mu,\nu\in\left\{ A,B,C,D \right\}}$ label the flavors. The nearest-neighbour coupling $J_{\vec{\imath},\vec{\jmath}}$ is given by
\begin{equation}
  \label{eq:multiboson-Hxy-Jxy}
  J_{\vec{\imath},\vec{\jmath}} =
  \begin{cases}
    J_x & \text{for intra-chain bonds,} \\
    J_y & \text{for inter-chain bonds.} \\
  \end{cases}
\end{equation}
At the isotropic point $J_{x} = J_{y}$, the model describes a square lattice whereas the regime $J_{y}/J_{x} = 0$ corresponds to decoupled chains. The states of the model of interest are the states of the six-dimensional fully antisymmetric self-adjoint representation. We will assume a N\'eel-type ordering with a bipartite configuration, where we have the flavors $A,B$ on one sublattice and the flavors $C,D$ on the other sublattice. 
Assuming the existence of such a magnetic phase, we will apply the multiboson approach \cite{masashige_transverse_2010,romhanyi_multiboson_2012,penc_spin-stretching_2012,kim_linear_2017} to study the behavior of the ordered magnetic moment of the system as a function of the inter-chain coupling $J_{y}$ in the linear flavor-wave approximation.
Within this approach, a boson is attributed to each of the six existing states in the irreducible representation. 
We will thus be working in terms of the composite particles, not in terms of the individual flavor particles.

\subsection{\label{sec:lfwt-multiboson-H}The LFWT multiboson Hamiltonian}

Let the six states of the antisymmetric irrep $\tiny{\yng(1,1)}$ be
\begin{equation}
  \begin{aligned}
    \label{eq:lfwt-su4-basis}
    \overline{AB} =&\frac{\ket{AB}-\ket{BA}}{\sqrt{2}}, &\overline{AC} =& \frac{\ket{AC}-\ket{CA}}{\sqrt{2}},\\
    \overline{DA} =&\frac{\ket{DA}-\ket{AD}}{\sqrt{2}}, &\overline{BC} =& \frac{\ket{BC}-\ket{CB}}{\sqrt{2}},\\
    \overline{BD} =&\frac{\ket{BD}-\ket{DB}}{\sqrt{2}}, &\overline{CD} =& \frac{\ket{CD}-\ket{DC}}{\sqrt{2}}.
  \end{aligned}
\end{equation}
The bar over the flavors is used as a reminder that the flavor indices are antisymmetric. We group these states into the set $\Gamma$:
\begin{equation}
  \Gamma = \{\overline{AB},\overline{AC},\overline{DA},\overline{BC},\overline{BD},\overline{CD}\}.
\end{equation}
\begin{figure}[h]
  \centerline{\hspace{1.6cm}\includegraphics[width=7.1cm]{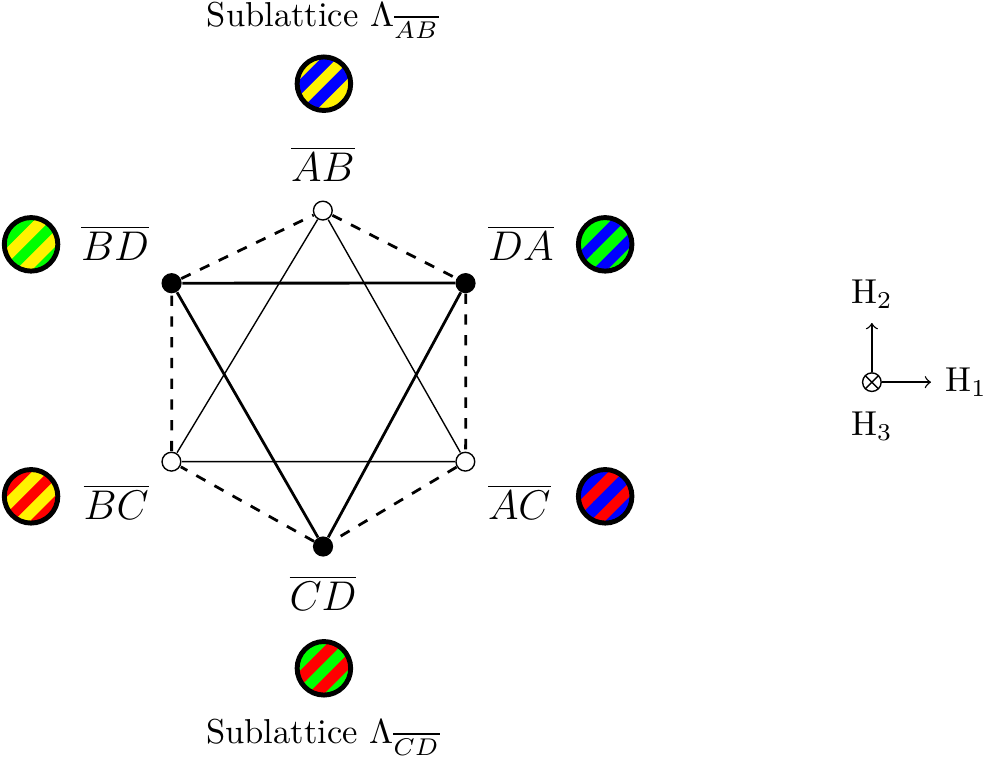}}
  \caption{\label{fig:weight-diag}Weight diagram of the six-dimensional antisymmetric $\mathrm{SU}(4)$ irrep. The flavors $A,B,C,D$ are represented by the colors blue, yellow, red, green respectively, as in Fig~\ref{fig:lattice}. The weight diagram  is in three dimensions in the Cartan-Weyl basis $\{\mathrm{H}_{1} \propto \hat S^A_A- \hat S^B_B, \mathrm{H}_{2}\propto \hat S^A_A+  \hat S^B_B-2  \hat S^C_C, \mathrm{H}_{3}\propto  \hat S^A_A+ \hat S^B_B+ \hat S^C_C-3 \hat S^D_D\}$, since $\mathrm{SU}(4)$ is a group of rank 3, and its vertices form an octahedron. The plane with circles is beneath the plane with dots. The states $\overline{AB}$ and $\overline{CD}$ are antipodal    and are farthest from each other.}
\end{figure}
The states are represented in the weight diagram in Fig.~\ref{fig:weight-diag}. Let us attribute a boson to each of these states. In other words, the bosons $d^{}_{\overline{AB}},d^{}_{\overline{AC}},d^{}_{\overline{DA}},d^{}_{\overline{BC}},d^{}_{\overline{BD}},d^{}_{\overline{CD}}$ and their adjoint counterparts will be used to create and annihilate the six states of the irrep. Since our model has one composite particle per site, we have the constraint
\begin{equation}
  \label{eq:lfwt-6-constraint}
  \sum\limits_{\eta \in \Gamma} d^{\dagger}_{\eta} d^{}_{\eta} = n_{c},
\end{equation}
with $n_{c}=1$ for each site, where the boson index ${\eta \in \Gamma}$ refers to the individual states in $\Gamma$. As for the $\mathrm{SU}(4)$ generators $\hat{S}^{\mu}_{\nu}(i)$ on a site $i$, they can be written as
\begin{equation}
  \label{eq:lfwt-6-Smn}
  \hat{S} ^{\mu}_{\nu}(i) = \sum\limits_{\substack{\alpha=A\\\alpha \neq \mu,\nu}}^{D}
  d^{\dagger}_{\overline{\alpha\nu}}(i) d^{}_{\overline{\alpha\mu}}(i) - \delta_{\mu,\nu} \frac{1}{2} n_{c},
\end{equation}
where the (antisymmetric) indices of the bosons are now ordered in such a way that they correspond to the labels of the states. When reordering the indices, the sign of the permutations needs to be taken into account, i.e., $d^{\dagger}_{\overline{\nu\mu}}=-d^{\dagger}_{\overline{\mu\nu}}$, to reflect the antisymmetry of the states of this irrep. As an example, the generator $\hat{S} ^A_B$ is given by
\begin{equation}
  \label{eq:-lfwt-Sab-ex}
  \hat{S} ^{A}_{B} = d^{\dagger}_{\overline{BC}} d^{}_{\overline{AC}} - d^{\dagger}_{\overline{BD}} d^{}_{\overline{DA}}.
\end{equation}
These expressions of $\hat{S}^{\mu}_{\nu}$ in terms of the bosons $d$ is a valid bosonic representation of the $\mathrm{SU}(4)$ generators as it satisfies the $\mathrm{SU}(N)$ commutation relation
\begin{equation}
  \label{eq:lfwt-sun-comm}
  \left[ \hat{S}^{\alpha}_{\beta}, \hat{S}^{\mu}_{\nu} \right] = \delta ^{\alpha}_{\nu} \hat{S}^{\mu}_{\beta} - \delta ^{\mu}_{\beta} \hat{S}^{\alpha}_{\nu}.
\end{equation}
Without loss of generality, the classical ground-state N\'eel order can be assumed to be composed of the state $\overline{AB}$ on sublattice $\Lambda_{\overline{AB}}$ and the state $\overline{CD}$ on sublattice $\Lambda_{\overline{CD}}$. Note that these two states are the farthest apart from one another in the weight diagram in Fig.~\ref{fig:weight-diag}. With the assumption of small fluctuations around our flavor order, the Holstein-Primakoff prescription can be used by considering the limit $n_{c} \rightarrow \infty$. Let us introduce the pair of Holstein-Primakoff bosons $a^{\phantom{\dagger}}_\eta(i)$  and $a^{\dagger}_\eta(i)$ for the sublattice $\Lambda_{\overline{AB}}$. Using the constraint~\eqref{eq:lfwt-6-constraint}, the prescription leads to the following equations:
\begin{subequations}
  \begin{align}
    \label{eq:lfwt-6-hp_a}
    d^{\dagger}_{\overline{AB}}(i), d^{}_{\overline{AB}}(i) &\rightarrow \sqrt{n_{c} - \sum\limits_{\eta } a^{\dagger}_{\eta}(i) a^{}_{\eta}}(i), \\
    d^{\dagger}_{\eta}(i) &\rightarrow a^{\dagger}_{\eta}(i),  \\
    d^{\phantom{\dagger}}_{\eta}(i) &\rightarrow a^{\phantom{\dagger}}_{\eta}(i),
  \end{align}
\end{subequations}
where the index $\eta$ labels the non-condensed bosons in the set $\Gamma$, {\it i.e.} $\eta \in \Gamma \setminus \{\overline{AB}\} = \Gamma_{\overline{AB}}$, with
\begin{equation}
  \Gamma_{\overline{AB}}  := \{\overline{AC},\overline{DA},\overline{BC},\overline{BD},\overline{CD}\}.
\end{equation}
Similarly, we can introduce the Holstein-Primakoff bosons $b^{\phantom{\dagger}}_\eta(j)$  and $b^{\dagger}_\eta(j)$ for the other sublattice $\Lambda_{\overline{CD}}$. In this case, we get
\begin{subequations}
  \begin{align}
    \label{eq:lfwt-6-hp_b}
    d^{\dagger}_{\overline{CD}}(j), d^{}_{\overline{CD}}(j) &\rightarrow \sqrt{n_{c} - \sum\limits_{\eta } b^{\dagger}_{\eta}(j) b^{}_{\eta}}(j), \\
    d^{\dagger}_{\eta}(j) &\rightarrow b^{\dagger}_{\eta}(j),  \\
    d^{\phantom{\dagger}}_{\eta}(j) &\rightarrow b^{\phantom{\dagger}}_{\eta}(j),
  \end{align}
\end{subequations}
with $\eta \in \Gamma \setminus \{\overline{CD}\} = \Gamma_{\overline{CD}}$. 

The square root expansion in $1/n_{c}$ yields the quadratic Hamiltonian $\mathcal{H}^{(2)}$. We can perform the Fourier transform on $\mathcal{H}^{(2)}$ with
\begin{equation}
  \label{eq:multiboson-su4-sq-FT}
  \begin{aligned}
    a^{}_{\eta}(i) = \sqrt{\frac{2}{N_{s}}} \sum\limits_{\mathbf{k}\in \text{RBZ}} a^{}_{\eta}(\mathbf{k})e^{-i \mathbf{k} \cdot \mathbf{r}_{i}},\\
    b^{}_{\eta}(j) = \sqrt{\frac{2}{N_{s}}} \sum\limits_{\mathbf{k}\in \text{RBZ}} b^{}_{\eta}(\mathbf{k})e^{-i \mathbf{k} \cdot \mathbf{r}_j},
  \end{aligned}
\end{equation}
in which $\eta \in \Gamma$ is the state index, $N_{s}$ is the number of sites and the sums run over the reduced magnetic Brillouin zone. The Fourier-transformed quadratic Hamiltonian $\mathcal{H}^{(2)}$ is finally given by
\begin{equation}
  \label{eq:lfwt-6-H2}
  \mathcal{H}^{(2)} = \mathcal{H}^{(2)}_{0} + \mathcal{H}^{(2)}_{1} + \mathcal{H}^{(2)}_{2} + \mathcal{H}^{(2)}_{3}
  + \mathcal{H}^{(2)}_{4}
\end{equation}
where
\begin{equation}
  \label{eq:lfwt-6-H2-sub}
  \begin{aligned}
    \mathcal{H}^{(2)}_{0} = n_{c} & \sum _{\mathbf{k}\in \text{RBZ}} 2 \mathcal{A} \left[ a_{\overline{CD},\mathbf{k}}^{\dagger}\, a^{}_{\overline{CD},\mathbf{k}} + b^{\dagger}_{\overline{AB},\mathbf{k}}\, b^{}_{\overline{AB},\mathbf{k}} \right], \\
    \mathcal{H}^{(2)}_{1} = n_{c} & \sum _{\mathbf{k}\in \text{RBZ}} \left[ \mathcal{A} \left( a_{\overline{AC},\mathbf{k}}^{\dagger}\,a_{\overline{AC},\mathbf{k}}^{} + b_{\overline{BD},\mathbf{k}}^{\dagger}\,b_{\overline{BD}} \right) \right. \\
    +& \left. \mathcal{B}_{\mathbf{k}} \left(a_{\overline{AC},\mathbf{k}}^{\dagger}\, b^{\dagger}_{\overline{BD},-\mathbf{k}} + a_{\overline{AC},\mathbf{k}}^{}\,b_{\overline{BD},-\mathbf{k}}^{} \right) \right], \\
    \mathcal{H}^{(2)}_{2} = n_{c} & \sum _{\mathbf{k}\in \text{RBZ}} \left[ \mathcal{A} \left( a_{\overline{BD},\mathbf{k}}^{\dagger}\,a_{\overline{BD},\mathbf{k}}^{} + b_{\overline{AC},\mathbf{k}}^{\dagger}\,b_{\overline{AC},\mathbf{k}}^{} \right) \right.\\
    +& \left. \mathcal{B}_{\mathbf{k}} \left( a_{\overline{BD},\mathbf{k}}^{\dagger}\, b_{\overline{AC},-\mathbf{k}}^{\dagger} +a_{\overline{BD},\mathbf{k}}^{}\, b_{\overline{AC},-\mathbf{k}}^{} \right) \right], \\
    \mathcal{H}^{(2)}_{3} = n_{c} & \sum _{\mathbf{k}\in \text{RBZ}} \left[ \mathcal{A} \left( a_{\overline{DA},\mathbf{k}}^{\dagger}\,a_{\overline{DA},\mathbf{k}}^{}+ b_{\overline{BC},\mathbf{k}}^{\dagger}\,b_{\overline{BC},\mathbf{k}}^{} \right) \right. \\
    +& \left. \mathcal{B}_{\mathbf{k}} \left( a_{\overline{DA},\mathbf{k}}^{\dagger}\,b_{\overline{BC},-\mathbf{k}}^{\dagger} +
        a_{\overline{DA},\mathbf{k}}^{}\, b_{\overline{BC},-\mathbf{k}}^{} \right) \right], \\
    \mathcal{H}^{(2)}_{4} = n_{c} & \sum _{\mathbf{k}\in \text{RBZ}} \left[ \mathcal{A} \left(
        a_{\overline{BC},\mathbf{k}}^{\dagger}\, a_{\overline{BC},\mathbf{k}}^{} + b_{\overline{DA},\mathbf{k}}^{\dagger}\,b_{\overline{DA},\mathbf{k}}^{} \right)
    \right.\\
    +& \left. \mathcal{B}_{\mathbf{k}} \left( a_{\overline{BC},\mathbf{k}}^{\dagger}\,b_{\overline{DA},-\mathbf{k}}^{\dagger} + a_{\overline{BC},\mathbf{k}}^{}\,b_{\overline{DA},-\mathbf{k}}^{} \right) \right],
  \end{aligned}
\end{equation}
with
\begin{equation}
  \label{eq:lfwt-6-ab}
  \begin{aligned}
    \mathcal{A} &= 2 J_{x}+2 J_{y},\\
    \mathcal{B}_{\mathbf{k}} &= 2 J_{x} \cos k_{x} +2 J_{y} \cos k_{y}.
  \end{aligned}
\end{equation}
Note that all the terms in $\mathcal{H}^{(2)}$ are of the same order in our expansion parameter $n_{c}$. Let us reestablish the constraint~\eqref{eq:lfwt-6-constraint} by setting $n_c=1$. The terms $\mathcal{H}^{(2)}_{1,\dots,4}$ in Eq.~\eqref{eq:lfwt-6-H2} can be diagonalized separately with the Bogoliubov transformation in an identical fashion. For instance, the diagonalization of bosons $a_{\overline{AC},\mathbf{k}}, b_{\overline{BD},\mathbf{k}}$ in $\mathcal{H}^{(2)}_{1}$ can be performed with
\begin{equation}
  \label{lfwt-6-bogoliubov}
  \begin{pmatrix}
    \ba^{\dagger}_{\overline{AC},\mathbf{k}}\\
    \bb^{}_{\overline{BD},-\mathbf{k}}\\
  \end{pmatrix}
  =
  \begin{pmatrix}
    u_{\mathbf{k}} & v_{\mathbf{k}}\\
    v_{\mathbf{k}} & u_{\mathbf{k}}\\
  \end{pmatrix}
  \begin{pmatrix}
    a^{\dagger}_{\overline{AC},\mathbf{k}}\\
    b^{}_{\overline{BD},-\mathbf{k}}\\
  \end{pmatrix},
\end{equation}
where
\begin{equation}
  u_{\mathbf{k}} = \sqrt{\frac{1}{2} \left( \frac{\mathcal{A}}{\omega_{\mathbf{k}}} + 1 \right)}, \quad v_{\mathbf{k}} = \sqrt{\frac{1}{2}\left( \frac{\mathcal{A}}{\omega_{\mathbf{k}}} - 1 \right)}.  
\end{equation}
Hence, the diagonalized quadratic Hamiltonian finally reads as
\begin{widetext}
  \begin{equation}
    \label{eq:lfwt-6-H-final}
    \mathcal{H}^{(2)} = \sum\limits_{\mathbf{k} \in \text{RBZ}} \left\{
      \sum\limits_{\substack{\eta \in \Gamma'}} \left[
        \omega_{\mathbf{k}} \left( \ba^{\dagger}_{\eta,\mathbf{k}} \ba^{}_{\eta,\mathbf{k}} + \frac{1}{2} \right)
        + \omega_{\mathbf{k}} \left( \bb^{\dagger}_{\eta,\mathbf{k}} \bb^{}_{\eta,\mathbf{k}} + \frac{1}{2} \right) \right]
      + 4(J_x+J_y) \left( a^{\dagger}_{\overline{CD},\mathbf{k}} a^{}_{\overline{CD},\mathbf{k}} + b^{\dagger}_{\overline{AB},\mathbf{k}}
        b^{}_{\overline{AB},\mathbf{k}} \right)  \right\} + \text{const.},
  \end{equation}
\end{widetext}
where $\Gamma' = \Gamma_{\overline{AB}}\cap \Gamma_{\overline{CD}} = \{\overline{AC},\overline{DA},\overline{BC},\overline{BD}\}$ and 
\begin{equation}
  \begin{aligned}
    \label{eq:lfwt-w}
    \omega_{\mathbf{k}} &= \sqrt{\mathcal{A}^{2} - \mathcal{B}_{\mathbf{k}}^{2}}\\
    &=  \sqrt{(2 J_x + 2 J_y)^2 - (2 J_{x} \cos k_{x} + 2 J_{y} \cos k_{y})^{2}}.
  \end{aligned}
\end{equation}

There are eight dispersive modes and two localized modes. The flat localized modes stem from multipolar transitions requiring more than one flavor exchange, and, thus, these excitations do not interact in the quadratic order of our expansion in $n_{c}$ \cite{romhanyi_multiboson_2012,kim_linear_2017}. Moreover, in the subsequent calculation, it can be seen that they do not contribute to the ordered moment in the harmonic approximation. It is also worthwhile noting that when applying the LFWT calculations using a different boson representation, namely the Read and Sachdev bosonic representation \cite{Read89}, it can be shown that only the eight dispersive modes are present in the harmonic approximation \cite{kim_linear_2017}.

\subsection{\label{sec:lfwt-6-magnetization}Magnetization and the dimensional crossover}
\begin{figure}[t]
  \centerline{\includegraphics[width=8cm]{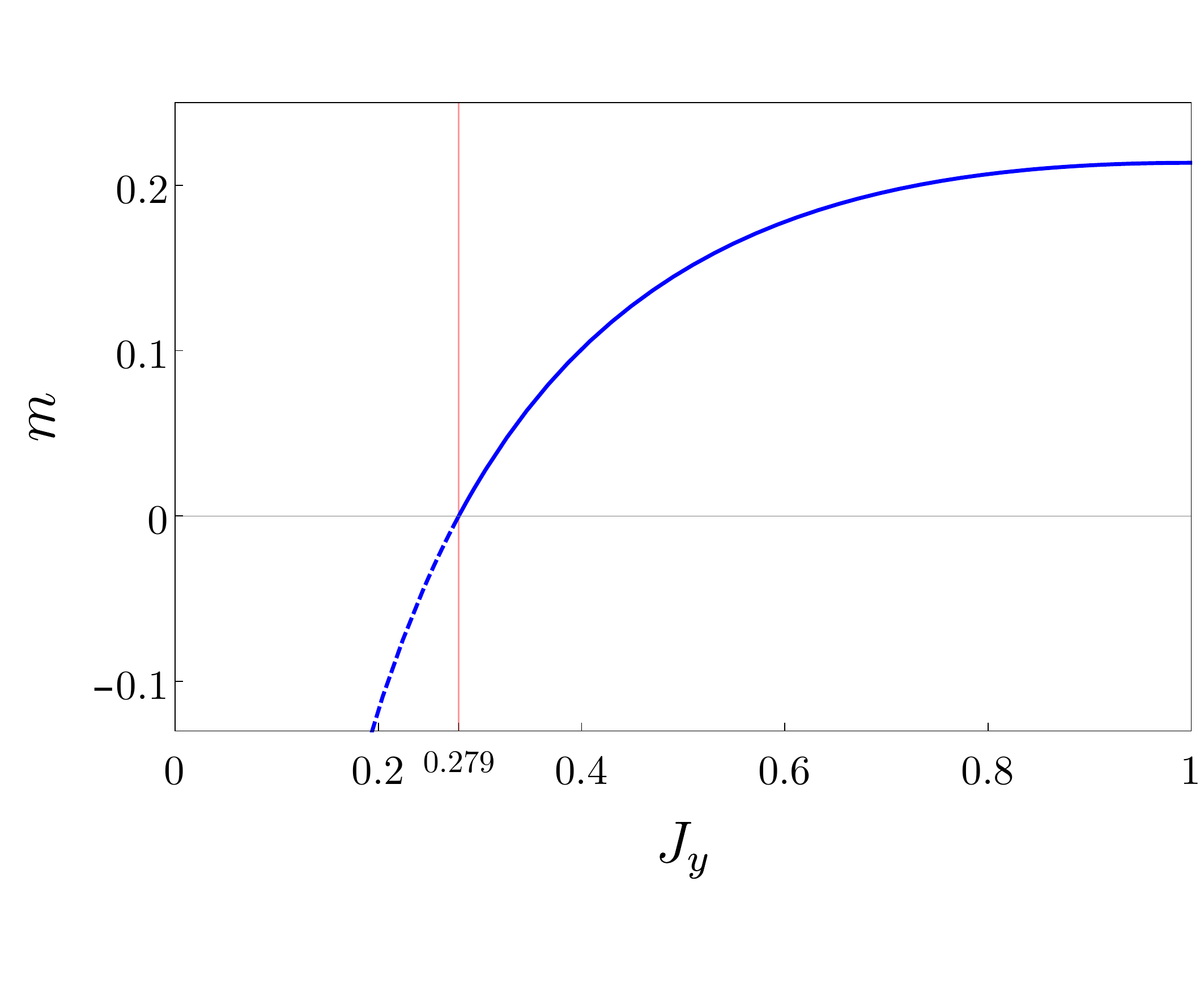}}
  \caption{\label{fig:lfwt-mag} Magnetization calculated from LFWT, as a function of $J_{y}$, while keeping $J_x=1$. The magnetization below $J_{y}^{c} = 0.279$ is negative, suggesting that the order is completely destroyed below this value.}
\end{figure}

Let us now study the magnetization and the dimensional crossover of the system. For a bipartite lattice with $\mathrm{SU}(4)$ flavors $A,B$ and $C,D$ on the two different sublattices, the ordered magnetic moment on site $i \in \Lambda_{\overline{AB}}$ can be defined as
\begin{equation}
  \label{eq:lfwt-mag}
  m_i = \frac{1}{2}\left\langle  \hat{S}^{A}_{A}(i) + \hat{S}^{B}_{B}(i) - \hat{S}^{C}_{C}(i) - \hat{S}^{D}_{D}(i) \right\rangle,
\end{equation}
such that the classical Néel configuration yields $m_{i}=n_{c}$ and the disordered case yields $m_{i}=0$. Using Eq.~\eqref{eq:lfwt-6-Smn}, this becomes
\begin{align}
  \label{eq:lfwt-mag-1}
  m_{i}(J_x,J_y) &= n_{c} 
                   - \left\langle \hat{n}_{\overline{AC}}(i) \right\rangle 
                   - \left\langle \hat{n}_{\overline{AD}}(i) \right\rangle 
                   - \left\langle \hat{n}_{\overline{BC}}(i) \right\rangle \nonumber\\
                 &\phantom{=} 
                   - \left\langle \hat{n}_{\overline{BD}}(i) \right\rangle 
                   - 2 \left\langle \hat{n}_{\overline{CD}}(i) \right\rangle,
\end{align}
where $\hat{n}_{\eta} = a ^{\dagger}_{\eta}(i) a ^{}_{\eta}(i)$ with $\eta \in \Gamma_{\overline{AB}}$.  Within the LFWT and with our constraint $n_{c}=1$, the ordered moment $m_{i}$ is finally given by
\begin{equation}
  \label{eq:lfwt-6-mag-i}
  m_{i}(J_x,J_y) 
  = 1 - 4 \left\langle v^{2}_{\mathbf{k}} \right\rangle,
\end{equation}
where we used the fact that $\left\langle \hat{n}_{\overline{CD}} \right\rangle = 0$ in the harmonic approximation, {\it i.e.} the localized bands do not contribute to the reduction of the magnetization. The ordered moment on the sublattice $\Lambda_{\overline{CD}}$ can be defined to be $m_{j}=-m_{i}$, and we thus consider $m=m_{i}$ only in the following and fix the value of the intra-chain coupling $J_x = 1$ for simplicity. In the isotropic case, $J_y = 1$, it has already been concluded in Ref.~[\onlinecite{kim_linear_2017}] that the magnetic moment retains a finite value,
\begin{equation}
  m= 0.214\;.
\end{equation} 
Though the correction to the magnetization $1 - m = 0.786$ is rather large, this would suggest a potential flavor order of the system. From this isotropic point, we can now investigate how the value of ordered moment decreases as we decrease the $J_y$.  As the magnetization $m$ vanishes when
\begin{equation}
  \label{eq:4}
  J_{y}^{\mathrm{c}} = 0.279\;,
\end{equation}
we can conclude that the dimensional crossover we are searching for happens at this point. Below this value of $J_{y}$, quantum fluctuations completely destroy the flavor order, indicating a possible phase transition. The ordered moment is plotted in Fig~\ref{fig:lfwt-mag}.
The LFWT thus predicts a phase transition from the N\'eel ordered state.

\section{\label{sec:QMC}  Auxiliary field quantum Monte Carlo }
\begin{figure}[t]
\centering
\includegraphics[width=0.85\linewidth]{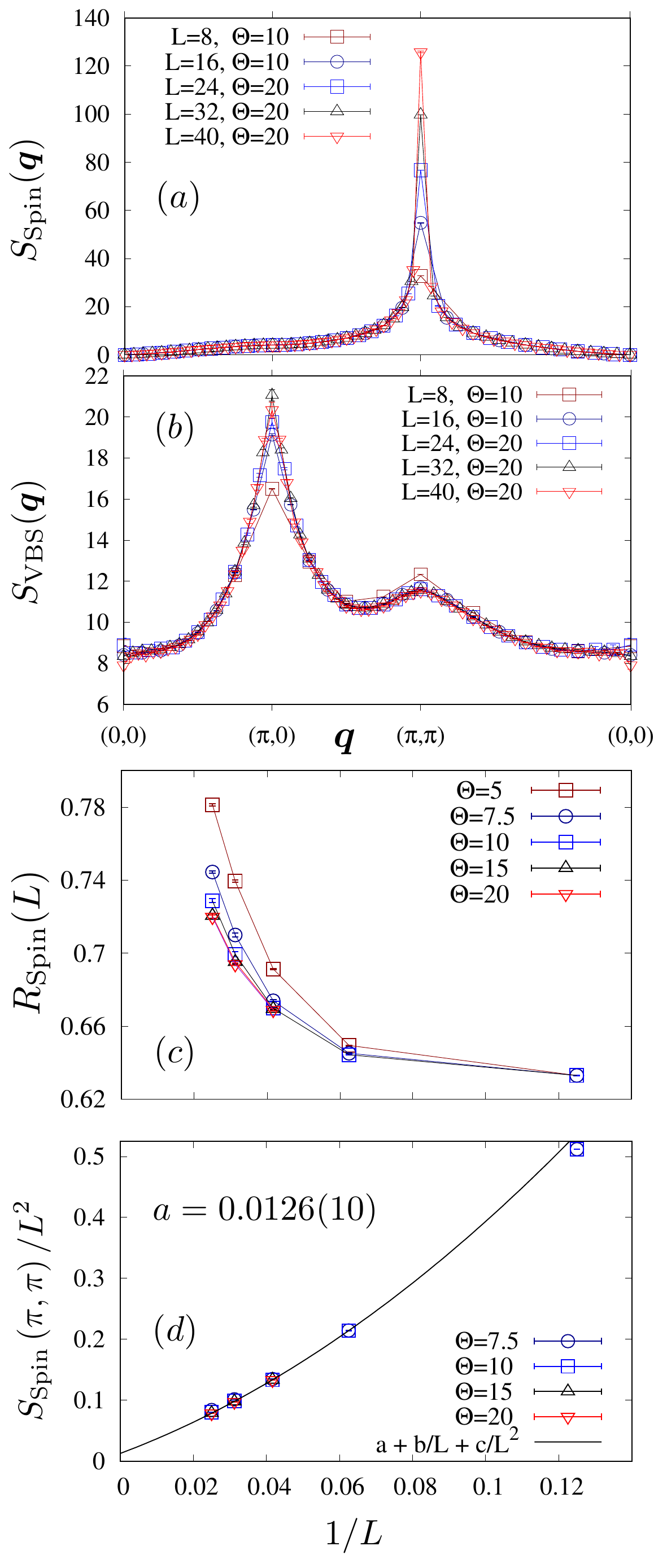}
\caption{QMC results for the isotropic SU(4) Heisenberg model, $J_x=J_y=1$. 
(a) 
Spin-Spin and (b) VBS correlation functions $S_{\text{VBS}}(\mathbf{q}) = \sum_{\delta} \left[ S_{\text{VBS}}  (\mathbf{q})\right]_{\delta,\delta} $  for various lattice sizes along a high symmetry path in the  Brillouin zone.    The projection parameter $\Theta$ grows as a function of system size $L$ so to guarantee that  we have indeed converged to the ground state. 
(c)  Spin
  correlation ratio. For each system size we have checked convergence in the projection parameter.  This quantity grows
  but shows no clear saturation to unity for lattice sizes up to  $40 \times 40$.   The data are consistent with a small
  local moment.
(d) Spin-spin correlation  at the antiferromagnetic wave vector
  divided by the volume, the ordered moment corresponds to $m^2 = S_{\text{Spin}}(\mathbf{Q})/L^2$.    For each system size we have checked for
  convergence in $\Theta$.  Extrapolation of converged results support a small but finite local moment m.}  \label{Scale_spin.fig}
\end{figure}

Auxiliary field QMC simulations of the SU($N$) Hubbard model, Eq.~(\ref{eq:intro-hubbard}),   demonstrate  that charge fluctuations   favor magnetic ordering  at half-filling for  even values of $N$ \cite{Wang14b}.     At $N=6$  one observes a VBS state in the Heisenberg limit \cite{assaad_phase_2005},  $U /t \rightarrow \infty$, and a magnetically ordered state  below a  critical value of $U/t$.       For the strong coupling $N=4$  model,   the magnetic moment is a  decreasing  function of $U/t$  and in the Heisenberg limit it is to date not clear if the ground state is magnetically ordered.  Our first aim is to  carry out  more precise simulations  than in Ref.~[\onlinecite{assaad_phase_2005}] of the SU(4) Heisenberg model.  We will see that the model has a small but finite magnetic moment.  Having established  order in the isotropic case,  we will then search for the signatures of the dimensional crossover in the spin and VBS correlation functions.

\subsection{Auxiliary field QMC: formulation.}
It is beyond the scope of this article to provide a full description of   the auxiliary field QMC approach. Here we will restrict ourselves to  a formulation of the model akin to be implemented in the  Algorithms for Lattice Fermions (ALF) library  \cite{ALF_v1}.     Our starting point  is a fermionic representation of the  SU($N$) generators, 
\begin{equation}
\hat{S} ^{\mu}_{\nu}(i)    = \hat{c}^{\dagger}_{i,\nu} \hat{c}^{}_{i,\mu}   - \frac{\delta_{\mu,\nu}}{N}  \sum_{\alpha=1}^{N} \hat{c}^{\dagger}_{i,\alpha} \hat{c}^{}_{i,\alpha} \;,
\label{eq:SmunuFA}
\end{equation}
where the  half-filling constraint 
\begin{equation}
  N_c = \sum_{\nu=1}^{N}\hat{c}^{\dagger}_{i,\nu} \hat{c}^{}_{i,\nu}   = \frac{N}{2}
\end{equation}
selects  the fully antisymmetric self-adjoint representation.   In this representation, the Heisenberg model,  $
  \mathcal{H} =  \sum\limits_{\langle i,j \rangle}   J_{\langle i,j \rangle }  \sum\limits_{\mu,\nu} \hat{S} ^{\mu}_{\nu}(i) \hat{S}
  ^{\nu}_{\mu}(j)$ 
  reads 
\begin{equation}
\begin{aligned}
\mathcal{H}   & = \mathcal{H}_J + \mathcal{H}_U \;,\\
\mathcal{H}_J & =  -\frac{1}{2 }  \sum_{ \langle i,j \rangle  }  J_{\langle i,j \rangle } \left(
           \hat{D}^{\dagger}_{ i,j }\hat{D}^{\phantom\dagger}_{ i j }  +
            \hat{D}^{\phantom\dagger}_{ i, j } \hat{D}^{\dagger}_{ i,j }  \right) \;, \\
\mathcal{H}_U & = U \sum_{i} \left(
            \sum_{\nu=1}^{N} \left[  \hat{c}^{\dagger}_{i,\nu}  \hat{c}^{\phantom\dagger}_{i,\nu} -  \frac{1}{2}  \right] \right)^2   \;.
\end{aligned}
\end{equation}

In the above, $ \hat{D}^{\dagger}_{ i,j }\ =  \sum_{\nu}  \hat{c}^{\dagger}_{i,\nu}  \hat{c}^{}_{j,\nu} $   and we have relaxed the  half-filled constraint   at the expense of the Hubbard interaction $\mathcal{H}_U $.   Since $ \left[ \mathcal{H}_U, \mathcal{H}_J \right]  =0 $ the  constraint will be automatically imposed  when carrying  out simulations at any finite positive value of $U$,  and in the  limit of infinite projection parameter  $\Theta$ (see below).  We use the equation: 
\begin{equation}
	\hat{D}^{\dagger}_{ i,j }\hat{D}^{\phantom\dagger}_{ i j }  +
            \hat{D}^{\phantom\dagger}_{ i, j } \hat{D}^{\dagger}_{ i,j }   = 
            \frac{1}{2} \left[  \left(\hat{D}^{\dagger}_{ i,j } + \hat{D}^{\phantom\dagger}_{ i j }  \right)^2  +  \left(i\hat{D}^{\dagger}_{ i,j } - i \hat{D}^{\phantom\dagger}_{ i j }  \right)^2  \right] 
\end{equation} 
so as to write the Hamiltonian in terms of perfect squares of  Hermitian operators as required by the standards of the ALF library \cite{ALF_v1}.    While ground state properties can be obtained using the grand canonical formulation of the auxiliary field QMC   and extrapolating to zero temperature, it is more convenient to adopt a projective scheme based on the  equation:
\begin{equation}
 \frac{\langle\psi_0|\hat{O}|\psi_0\rangle}{\langle\psi_0|\psi_0\rangle} = \lim_{\Theta \rightarrow \infty }   \frac{\langle\psi_T | e^{-\Theta \mathcal{H}/2} \hat{O} e^{-\Theta \mathcal{H}/2}  |\psi_T\rangle}
          {\langle\psi_T|e^{-\Theta \mathcal{H}} | \psi_T\rangle}
\end{equation}
provided that    $ \langle \psi_0 | \psi_T\rangle \neq 0 $. 
We have chosen the trial wave function to be the ground state of  the tight binding model on the  square lattice,
\begin{equation}
         \mathcal{H}_T =  - \sum_{ \langle i, j \rangle }   \sum_{\nu=1}^{N} \left(  \hat{c}^{\dagger}_{i,\nu}  \hat{c}^{\phantom{\dagger}}_{j,\nu} + \text{H.c.} \right),
\end{equation}
with anti-periodic (periodic) boundary conditions in the $x$ ($y$)  direction.     To study the dimensional crossover  we use the  exchange defined in  Eq. (\ref{eq:multiboson-Hxy-Jxy}), set $J_x$ to unity, the imaginary time step $\Delta \tau = 0.025$ and $U=0.25$.   For the considered values of the projection parameter, $\Theta$,  we have tested that this choice of the Hubbard interaction suffices to freeze the charge fluctuations  within  the statistical uncertainty. 

\begin{figure}[t]
\centering
\includegraphics[width=0.95\linewidth]{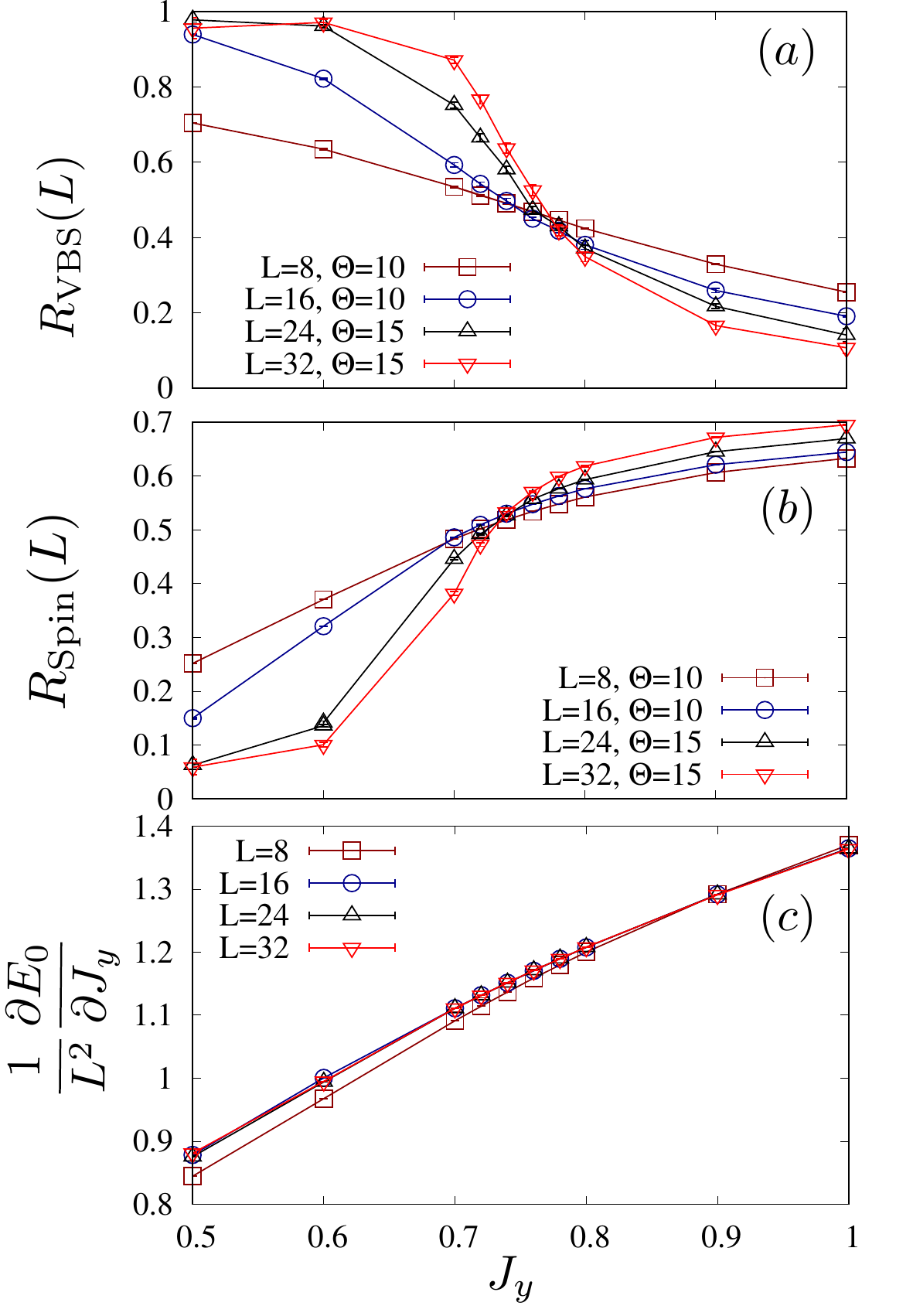}
\caption{(a) VBS  and (b) spin correlation ratios as a function of $J_y$, while keeping $J_{x}=1$.   The crossing in the spin and VBS channels are slightly shifted.  From the VBS data, one would have:  $J_y^{c} \simeq 0.76-0.78$  whereas for the spin $J_y^{c} \simeq 0.74-0.76$.   Given the overall scatter of the crossing point,  this difference is not significant enough to claim two separate transitions. (c)  $\frac{1}{L^2}\frac{ \partial E_0}{\partial J_y}$ shows no jump, thereby supporting a continuous transition. }
\label{Dim_crosover.fig}
\end{figure}

\subsection{Isotropic case}

To  pin down the nature of ground state of the SU(4) Heisenberg model,  we compute  equal time spin-spin correlation function
\begin{equation}
   S_{\text{Spin}}(\mathbf{q}) =  \frac{1}{L^2 } \sum_{i,j,\mu,\nu} e^{i \mathbf{q} \cdot (\mathbf{r}_{i}-\mathbf{r}_{j}) } \langle  \hat{S}^{\mu}_{\nu}(j)   \hat{S}^{\nu}_{ \mu}(i) \rangle \;.
\end{equation}
The $ S_{\text{Spin}}(\mathbf{q})$ fulfills the 
\begin{equation}
 \frac{1}{L^2} \sum_q S_{\text{Spin}}(\mathbf{q}) = C_1 
\end{equation}
sum rule, where $C_1$ is the value of the Casimir operator $\hat C_1 = \sum_{\mu,\nu}  \hat{S}^{\mu}_{\nu}(i)   \hat{S}^{\nu}_{ \mu}(i)$ on a site. For the 6-dimensional irreducible representation $C_1=5$.  The  above sum rule is valid only in the absence of charge fluctuations, so that it  provides an excellent crosscheck for the validity of our calculation and choice of the Hubbard U.  Indeed, our QMC calculations satisfied the sum rule up to $2.5\times 10^{-4}$ precision.  While in  Ref.~[\onlinecite{assaad_phase_2005}]   our biggest  size corresponded to   $24 \times 24$,   enhanced computer power allows us to reach ground state properties at $L=40$.

The results for $S_{\text{Spin}}(\mathbf{q})$  are shown in Fig.~\ref{Scale_spin.fig}(a) for different system sizes. We can observe a clear peak at $\mathbf{q} =  \mathbf{Q} = (\pi,\pi) $, which grows as the system size is increased, revealing the formation of the Ne\'el state.  To check the presence of long range order, we consider the correlation ratio: 
\begin{equation}
        R_{\text{Spin}}(L)  =  1 - \frac{S_{\text{Spin}}\left(\mathbf{Q} - \left( 0, 2 \pi/L \right) \right) } {S_{\text{Spin}}(\mathbf{Q}) }.
\end{equation}
This quantity scales to unity (zero) in the ordered  (disordered)  phase, and is a  renormalization group invariant quantity such that in the vicinity of a critical point --  where scaling holds -- we expect
\begin{equation}
        R_{\text{Spin}}(L)    =  F\left( (g-g_c)L^{1/\nu},L^{-\omega } \right).
\end{equation}
In the above, $g$ is the control  parameter, $\nu $ the correlation length exponent and $\omega$ the leading correction to  scaling exponent.   As apparent from Fig.~\ref{Scale_spin.fig}(c) the  ground state estimate of  $ R_{\text{Spin}}(L)  $  as a  function of system size is initially flat and then grows substantially when   $L \geq 24$.    This form of the correlation ratio strongly suggests that we are close to a critical point.  It is tempting to  interpret $N$  as a tuning parameter  that drives the system  from the N\'eel to VBS state. In this scenario  the local moment is small due to competing VBS fluctuations. To test this  we have computed the   VBS correlation functions: 
 \begin{equation}
 \begin{aligned}
         &    \left[ S_{\text{VBS}}  (\mathbf{q})\right]_{\delta,\delta' }   = 
             \frac{1}{L^2} \sum_{i,j}  e^{i \mathbf{q} \cdot \left( \mathbf{r}_i - \mathbf{r}_j \right) }  \times  \\   &  \left(
              \langle \hat{\Delta}_{i,i +\delta} \hat{\Delta}_{j,j+\delta'} \rangle  -
               \langle \hat{\Delta}_{i,i+\delta} \rangle \langle \hat{\Delta}_{j,j+\delta'} \rangle  \right)
\end{aligned}
\end{equation}
 with
\begin{equation*}
     \hat{\Delta}_{i,i+\delta}   =  \sum_{\mu,\nu}  \hat{S}^{\mu}_{\nu}(i)   \hat{S}^{\nu}_{\mu}(i + \delta).
\end{equation*}
Note that    to facilitate the calculation  of the dimer correlation function, we have used  $ \hat{S} ^{\mu}_{\nu}(i)    = \hat{c}^{\dagger}_{i,\nu} \hat{c}^{}_{i,\mu}   - \frac{1}{2}\delta_{\mu,\nu}   $  ({\it i.e.} the charge fluctuations in the diagonal part of $\hat{S} ^{\mu}_{\nu}(i)$, Eq.~(\ref{eq:SmunuFA}), are neglected). 
Figures~\ref{Scale_spin.fig}(a) and~\ref{Scale_spin.fig}(b)   plot  the  spin as well as VBS   correlation functions on our biggest lattice.   While the  antiferromagnetic  spin fluctuations  dominate, one observes strong
$\mathbf{q} = (0,\pi) $ and $\mathbf{q} = (\pi,0) $    VBS fluctuations thus lending support to the point of view that the SU(4)  quantum antiferromagnetic  is \textit{ close} to a quantum critical point. 

Finally, we calculate the value of the ordered moment.
In the pure Ne\'el state, where the fluctuations are fully neglected, $m=1$ and the correlations in real space are
\begin{equation}
   \sum_{\mu,\nu} \langle  \hat{S}^{\mu}_{\nu}(j)   \hat{S}^{\nu}_{ \mu}(i) \rangle
   = \begin{cases} 
     5 &\mbox{if } i=j; \\ 
     1 & \mbox{if $i\neq j$, same sublattice};  \\
     -1 & \mbox{if $i\neq j$, different sublattice}.
  \end{cases}
\end{equation}
Correspondingly, the correlation function in the reciprocal space, 
\begin{equation}
        S_{\text{Spin}}^{\text{Ne\'el}}(\mathbf{q}) = 4 + L^2 \delta_{\mathbf{q},\mathbf{Q}}  \;,
\end{equation}
shows a peak diverging with the system size at the ordering vector $\mathbf{Q}=(\pi,\pi)$. 

Figure~\ref{Scale_spin.fig}(d)   plots $S_{\text{Spin}}(\mathbf{Q})/L^2$ as a function of $1/L$ for the QMC calculation. The local moment, defined in Eq.~(\ref{eq:lfwt-mag}), corresponds to
\begin{equation}
 m^2 \equiv \lim_{L  \rightarrow \infty}   \frac{1}{L^2} S_{\text{Spin}}(\mathbf{Q}) \;.
\end{equation} 
A polynomial fit in $1/L$ using the values for $L=16,24,32,$ and $40$ gives $m^2 = 0.0126(10)  $.
   As apparent, large system sizes and  large projection parameters support a small but finite local moment in the thermodynamic limit.  In particular our  results  suggest that 
\begin{equation}
\label{m_QMC.Eq}
 m^{\text{QMC}}  =  0.11 \pm 0.04  
\end{equation}
and is hence two times smaller that the linear flavor-wave result.   As shown in the appendix,  this value of the local moment matches well  with the  one obtained from the Hubbard model in the large $U/t$ limit \cite{Wang14b}.

\subsection{Dimensional crosover}

To investigate the dimensional crossover,   we consider  again the spin and VBS correlation ratios.  As apparent in Figs.~\ref{Dim_crosover.fig} (a) and (b),  the data are consistent with a direct and continuous  transition between the  AFM and VBS at $J_y^c = 0.74 -  0.78$.    A more precise  study of the transition is certainly possible but  difficult.  In particular we have seen that   due to the small magnetic moment  of the AFM state in the isotropic limit, very large  system sizes are  required to merely establish  long range order.  Given  the numerically  accessible lattice sizes, we believe that these  difficulties  will hinder  an  accurate  estimate of the critical point as a function of dimensionality.      As mentioned  at the beginning of the section, charge fluctuations have the potential of enhancing the magnetic moment  in the isotropic limit, such that a model with charge fluctuations may be more suitable to   study the criticality of the dimensional crossover. 

Fig.~\ref{Dim_crosover.fig} (c)  plots 
\begin{equation}
	\frac{1}{L^2}  \frac{\partial E_0 } {\partial J_y} 
\end{equation}
as a function of $J_y$.  The smoothness of the function  constitutes an additional hint that the transition is continuous.


\section{\label{sec:conclusion}Conclusion}
Using QMC and LFWT, we investigated the $\mathrm{SU}(4)$ AFM Heisenberg model in the fully antisymmetric six-dimensional self-conjugate representation in two spatial dimensions and the dimensional crossover to one dimension. Both methods show that the isotropic model in 2D has AFM order, albeit with a very small magnetic moment according to the QMC data. The LFWT predicts a larger magnetic moment ($m= 0.214$) than the QMC calculations ($ m\simeq 0.11$). The dimensional crossover to 1D yields a phase transition from the N\'eel state to the VBS, and the critical value of the dimensional crossover is $J^{c}_{y} = 0.74 - 0.78$ according to QMC. The fading of the Néel phase during the dimensional crossover is also captured by the LFWT, although it overestimates the robustness of the N\'eel phase with a predicted transition value of $J^{c}_{y} = 0.279$.    We understand the  discrepancy between the  QMC and LFWT   calculations as a consequence of the Berry phase.     For the SU(2) model,  Haldane \cite{Haldane88}  has shown  that  skyrmion   changing configurations  (hedgehogs or monopoles) carry $C_4$    charge such that the proliferation of quadruple  monopole instances   leads to a  VBS state.  On the realm of the theory deconfined quantum criticality (DQC)  quadruple  monopole instances   are expected to be  irrelevant at criticality  and beyond criticality condense to form the  VBS state \cite{Senthil1490,Senthil04_1}.  Remarkably,  hedgehog  singularities and the conclusions of Ref.~[\onlinecite{Haldane88}]  can be  generalized to  SU($N$) \cite{Read89,Read90}.   LFWT  does not allow  for  singular field  configurations,  and the strong VBS fluctuations observed in the  QMC calculations suggest that they cannot be omitted  for an accurate description of the SU(4) quantum antiferromagnet.     In particular,   promoting $N$   to a continuous variable, our results show that the SU(4) quantum antiferromagnet is close   to a  putative  deconfined quantum critical  point  to the VBS.   Various,  yet to be numerically confirmed, field theories can be put forward to understand  this quantum  phase transition \cite{Read89,Xu18}  in a two dimensional setting.  Finally, the nature of the dimensional driven transition  to the VBS   remains to be studied.   In the realm of the theory of DQC, the reduction  of the lattice symmetry  from C$_4$ to C$_2$   allows for  double monopole instances in the   field theory.  A continuous transition -- as supported by the numerical data --  would require   double  monopole instances to be irrelevant at criticality. 

\begin{acknowledgments}
We thank  Jonas Schwab  and Congjun Wu  for  discussions. 
F.F.A. thanks  the DFG collaborative research center SFB1170 ToCoTronics (project C01)   for financial support as well as the W\"urzburg-Dresden Cluster of Excellence on Complexity and Topology in Quantum Matter -- \textit{ct.qmat} (EXC 2147, project-id 39085490). F.H.K. and F.M. acknowledge the financial support of the Swiss National Science Foundation. K.P. acknowledges the support of the Hungarian NKFIH Grant No. K 124176 and BME - Nanonotechnology and Materials Science FIKP grant of EMMI (BME FIKP-NAT). The authors gratefully acknowledge the Gauss Centre for Supercomputing e.V. (www.gauss-centre.eu) for providing computing time on the GCS Supercomputer SUPERMUC-NG at Leibniz Supercomputing Centre (www.lrz.de). K.P. is grateful for the hospitality of the EPFL where parts of the work have been completed.
\end{acknowledgments}

\appendix*
\section{Extrapolated value of $m$ from \textcite{Wang14b}}
\begin{figure}[h!]
\centering
\includegraphics[width=0.9\linewidth]{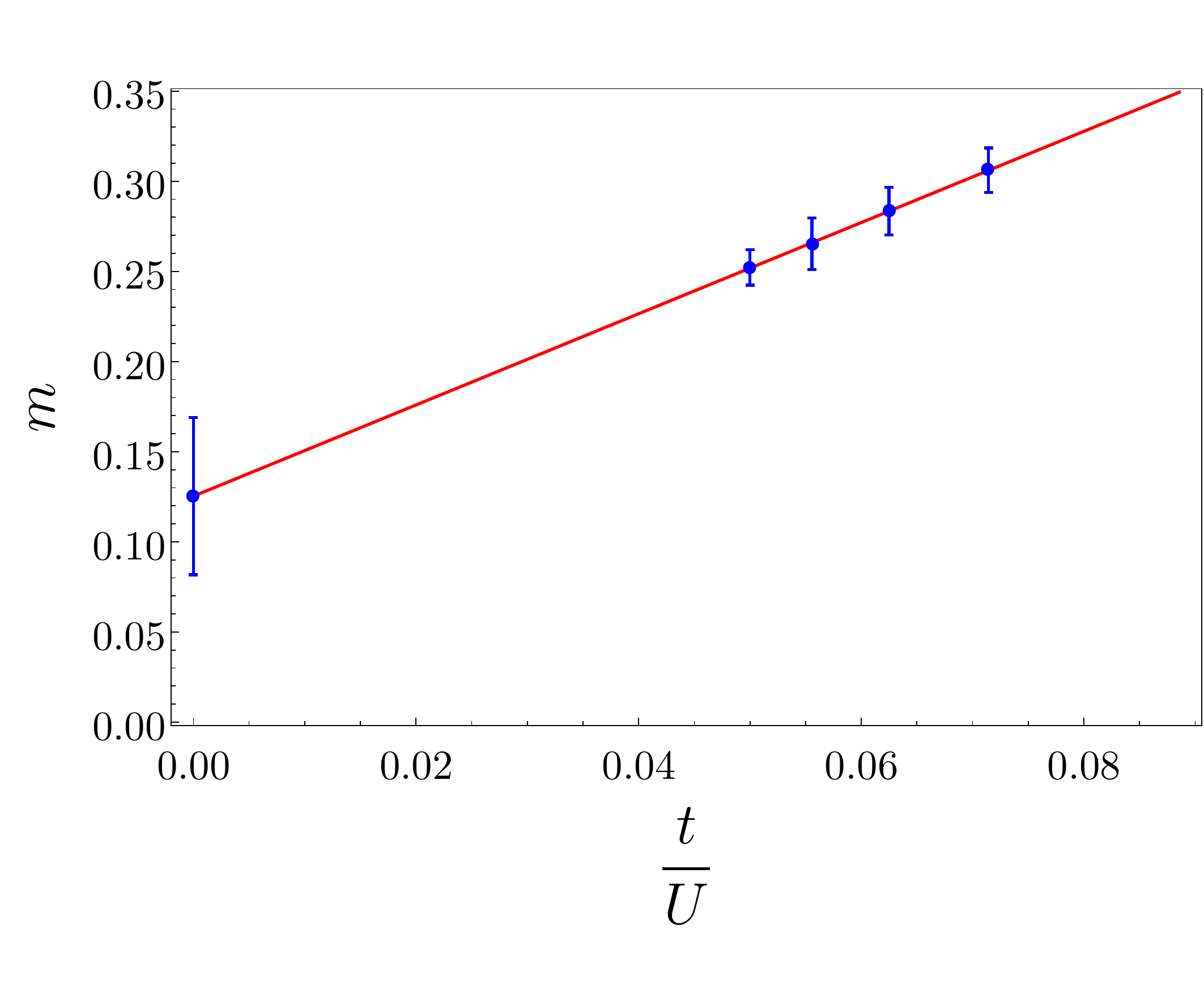}
\caption{\label{fig:m-Wu}Extrapolation of $m$ as a function of $\frac{t}{U}$ using data extracted from Fig. 4 of Ref.~[\onlinecite{Wang14b}]. In the limit $U/t \rightarrow \infty$, the linear fit in $t/U$ yields the magnetization value $m = 0.125 \pm 0.044$.   This extrapolated value compares favorably with  our estimate of Eq.~\ref{m_QMC.Eq}. }
\end{figure}

Ref.~[\onlinecite{Wang14b}] investigates the $\mathrm{SU}(4)$ Hubbard model in the 2D square lattice with the nearest-neighbor hopping integral $t$ and the on-site repulsion $U$. The pinning-field QMC method \cite{Assaad13}, which induces a symmetry-breaking, is used to probe the long-range magnetic order in the system.    In the absence of explicit symmetry breaking  we can only measure   spin-spin correlation functions and thereby determine the square of the local moment. As apparent in Fig.~\ref{Scale_spin.fig} (d)  very large systems are required so as  to   extract   reliably the value of the local moment.   For these specific cases, where the local moment is small, the pinning field approach seems superior. 
Fig. 4 in Ref.~\onlinecite{Wang14b} shows the magnetization $m$ as a function of $U/t$. Extracting the data from this figure, we have extrapolated the value of $m$ in the Heisenberg limit $U/t \rightarrow \infty$ using a linear fit in $t/U$, see Fig.~\ref{fig:m-Wu}. The obtained value, $m = 0.125 \pm 0.044$, is consistent with the value of $m$ (see Eq.~\ref{m_QMC.Eq}) obtained in this work  without introducing a symmetry breaking field.

\bibliographystyle{apsrev4-1}

\bibliography{fassaad,fkim}

\begin{thebibliography}{46}%
\makeatletter
\providecommand \@ifxundefined [1]{%
 \@ifx{#1\undefined}
}%
\providecommand \@ifnum [1]{%
 \ifnum #1\expandafter \@firstoftwo
 \else \expandafter \@secondoftwo
 \fi
}%
\providecommand \@ifx [1]{%
 \ifx #1\expandafter \@firstoftwo
 \else \expandafter \@secondoftwo
 \fi
}%
\providecommand \natexlab [1]{#1}%
\providecommand \enquote  [1]{``#1''}%
\providecommand \bibnamefont  [1]{#1}%
\providecommand \bibfnamefont [1]{#1}%
\providecommand \citenamefont [1]{#1}%
\providecommand \href@noop [0]{\@secondoftwo}%
\providecommand \href [0]{\begingroup \@sanitize@url \@href}%
\providecommand \@href[1]{\@@startlink{#1}\@@href}%
\providecommand \@@href[1]{\endgroup#1\@@endlink}%
\providecommand \@sanitize@url [0]{\catcode `\\12\catcode `\$12\catcode
  `\&12\catcode `\#12\catcode `\^12\catcode `\_12\catcode `\%12\relax}%
\providecommand \@@startlink[1]{}%
\providecommand \@@endlink[0]{}%
\providecommand \url  [0]{\begingroup\@sanitize@url \@url }%
\providecommand \@url [1]{\endgroup\@href {#1}{\urlprefix }}%
\providecommand \urlprefix  [0]{URL }%
\providecommand \Eprint [0]{\href }%
\providecommand \doibase [0]{http://dx.doi.org/}%
\providecommand \selectlanguage [0]{\@gobble}%
\providecommand \bibinfo  [0]{\@secondoftwo}%
\providecommand \bibfield  [0]{\@secondoftwo}%
\providecommand \translation [1]{[#1]}%
\providecommand \BibitemOpen [0]{}%
\providecommand \bibitemStop [0]{}%
\providecommand \bibitemNoStop [0]{.\EOS\space}%
\providecommand \EOS [0]{\spacefactor3000\relax}%
\providecommand \BibitemShut  [1]{\csname bibitem#1\endcsname}%
\let\auto@bib@innerbib\@empty
\bibitem [{\citenamefont {Wu}\ \emph {et~al.}(2003)\citenamefont {Wu},
  \citenamefont {Hu},\ and\ \citenamefont {Zhang}}]{wu_exact_2003}%
  \BibitemOpen
  \bibfield  {author} {\bibinfo {author} {\bibfnamefont {C.}~\bibnamefont
  {Wu}}, \bibinfo {author} {\bibfnamefont {J.-p.}\ \bibnamefont {Hu}}, \ and\
  \bibinfo {author} {\bibfnamefont {S.-c.}\ \bibnamefont {Zhang}},\ }\href
  {\doibase 10.1103/PhysRevLett.91.186402} {\bibfield  {journal} {\bibinfo
  {journal} {Phys. Rev. Lett.}\ }\textbf {\bibinfo {volume} {91}},\ \bibinfo
  {pages} {186402} (\bibinfo {year} {2003})}\BibitemShut {NoStop}%
\bibitem [{\citenamefont {Honerkamp}\ and\ \citenamefont
  {Hofstetter}(2004)}]{honerkamp_ultracold_2004}%
  \BibitemOpen
  \bibfield  {author} {\bibinfo {author} {\bibfnamefont {C.}~\bibnamefont
  {Honerkamp}}\ and\ \bibinfo {author} {\bibfnamefont {W.}~\bibnamefont
  {Hofstetter}},\ }\href {\doibase 10.1103/PhysRevLett.92.170403} {\bibfield
  {journal} {\bibinfo  {journal} {Phys. Rev. Lett.}\ }\textbf {\bibinfo
  {volume} {92}},\ \bibinfo {pages} {170403} (\bibinfo {year}
  {2004})}\BibitemShut {NoStop}%
\bibitem [{\citenamefont {Jaksch}\ and\ \citenamefont
  {Zoller}(2005)}]{jaksch_cold_2005}%
  \BibitemOpen
  \bibfield  {author} {\bibinfo {author} {\bibfnamefont {D.}~\bibnamefont
  {Jaksch}}\ and\ \bibinfo {author} {\bibfnamefont {P.}~\bibnamefont
  {Zoller}},\ }\href {\doibase 10.1016/j.aop.2004.09.010} {\bibfield  {journal}
  {\bibinfo  {journal} {Annals of Physics}\ }\textbf {\bibinfo {volume}
  {315}},\ \bibinfo {pages} {52 } (\bibinfo {year} {2005})},\ \bibinfo {note}
  {special Issue}\BibitemShut {NoStop}%
\bibitem [{\citenamefont {Cazalilla}\ \emph {et~al.}(2009)\citenamefont
  {Cazalilla}, \citenamefont {Ho},\ and\ \citenamefont
  {Ueda}}]{cazalilla_ultracold_2009}%
  \BibitemOpen
  \bibfield  {author} {\bibinfo {author} {\bibfnamefont {M.~A.}\ \bibnamefont
  {Cazalilla}}, \bibinfo {author} {\bibfnamefont {A.~F.}\ \bibnamefont {Ho}}, \
  and\ \bibinfo {author} {\bibfnamefont {M.}~\bibnamefont {Ueda}},\ }\href
  {\doibase 10.1088/1367-2630/11/10/103033} {\bibfield  {journal} {\bibinfo
  {journal} {New Journal of Physics}\ }\textbf {\bibinfo {volume} {11}},\
  \bibinfo {pages} {103033} (\bibinfo {year} {2009})}\BibitemShut {NoStop}%
\bibitem [{\citenamefont {Gorshkov}\ \emph {et~al.}(2010)\citenamefont
  {Gorshkov}, \citenamefont {Hermele}, \citenamefont {Gurarie}, \citenamefont
  {Xu}, \citenamefont {Julienne}, \citenamefont {Ye}, \citenamefont {Zoller},
  \citenamefont {Demler}, \citenamefont {Lukin},\ and\ \citenamefont
  {Rey}}]{gorshkov_two-orbital_2010}%
  \BibitemOpen
  \bibfield  {author} {\bibinfo {author} {\bibfnamefont {A.~V.}\ \bibnamefont
  {Gorshkov}}, \bibinfo {author} {\bibfnamefont {M.}~\bibnamefont {Hermele}},
  \bibinfo {author} {\bibfnamefont {V.}~\bibnamefont {Gurarie}}, \bibinfo
  {author} {\bibfnamefont {C.}~\bibnamefont {Xu}}, \bibinfo {author}
  {\bibfnamefont {P.~S.}\ \bibnamefont {Julienne}}, \bibinfo {author}
  {\bibfnamefont {J.}~\bibnamefont {Ye}}, \bibinfo {author} {\bibfnamefont
  {P.}~\bibnamefont {Zoller}}, \bibinfo {author} {\bibfnamefont
  {E.}~\bibnamefont {Demler}}, \bibinfo {author} {\bibfnamefont {M.~D.}\
  \bibnamefont {Lukin}}, \ and\ \bibinfo {author} {\bibfnamefont {A.~M.}\
  \bibnamefont {Rey}},\ }\href {\doibase 10.1038/nphys1535} {\bibfield
  {journal} {\bibinfo  {journal} {Nature Physics}\ }\textbf {\bibinfo {volume}
  {6}},\ \bibinfo {pages} {289} (\bibinfo {year} {2010})}\BibitemShut {NoStop}%
\bibitem [{\citenamefont {Taie}\ \emph {et~al.}(2012)\citenamefont {Taie},
  \citenamefont {Yamazaki}, \citenamefont {Sugawa},\ and\ \citenamefont
  {Takahashi}}]{taie_su6_2012}%
  \BibitemOpen
  \bibfield  {author} {\bibinfo {author} {\bibfnamefont {S.}~\bibnamefont
  {Taie}}, \bibinfo {author} {\bibfnamefont {R.}~\bibnamefont {Yamazaki}},
  \bibinfo {author} {\bibfnamefont {S.}~\bibnamefont {Sugawa}}, \ and\ \bibinfo
  {author} {\bibfnamefont {Y.}~\bibnamefont {Takahashi}},\ }\href {\doibase
  10.1038/nphys2430} {\bibfield  {journal} {\bibinfo  {journal} {Nature
  Physics}\ }\textbf {\bibinfo {volume} {8}},\ \bibinfo {pages} {825} (\bibinfo
  {year} {2012})}\BibitemShut {NoStop}%
\bibitem [{\citenamefont {Pagano}\ \emph {et~al.}(2014)\citenamefont {Pagano},
  \citenamefont {Mancini}, \citenamefont {Cappellini}, \citenamefont
  {Lombardi}, \citenamefont {Sch{\"a}fer}, \citenamefont {Hu}, \citenamefont
  {Liu}, \citenamefont {Catani}, \citenamefont {Sias}, \citenamefont
  {Inguscio},\ and\ \citenamefont {Fallani}}]{pagano_one-dimensional_2014}%
  \BibitemOpen
  \bibfield  {author} {\bibinfo {author} {\bibfnamefont {G.}~\bibnamefont
  {Pagano}}, \bibinfo {author} {\bibfnamefont {M.}~\bibnamefont {Mancini}},
  \bibinfo {author} {\bibfnamefont {G.}~\bibnamefont {Cappellini}}, \bibinfo
  {author} {\bibfnamefont {P.}~\bibnamefont {Lombardi}}, \bibinfo {author}
  {\bibfnamefont {F.}~\bibnamefont {Sch{\"a}fer}}, \bibinfo {author}
  {\bibfnamefont {H.}~\bibnamefont {Hu}}, \bibinfo {author} {\bibfnamefont
  {X.-J.}\ \bibnamefont {Liu}}, \bibinfo {author} {\bibfnamefont
  {J.}~\bibnamefont {Catani}}, \bibinfo {author} {\bibfnamefont
  {C.}~\bibnamefont {Sias}}, \bibinfo {author} {\bibfnamefont {M.}~\bibnamefont
  {Inguscio}}, \ and\ \bibinfo {author} {\bibfnamefont {L.}~\bibnamefont
  {Fallani}},\ }\href {\doibase 10.1038/nphys2878} {\bibfield  {journal}
  {\bibinfo  {journal} {Nature Physics}\ }\textbf {\bibinfo {volume} {10}},\
  \bibinfo {pages} {198} (\bibinfo {year} {2014})}\BibitemShut {NoStop}%
\bibitem [{\citenamefont {Scazza}\ \emph {et~al.}(2014)\citenamefont {Scazza},
  \citenamefont {Hofrichter}, \citenamefont {H{\"o}fer}, \citenamefont
  {De~Groot}, \citenamefont {Bloch},\ and\ \citenamefont
  {F{\"o}lling}}]{scazza_observation_2014}%
  \BibitemOpen
  \bibfield  {author} {\bibinfo {author} {\bibfnamefont {F.}~\bibnamefont
  {Scazza}}, \bibinfo {author} {\bibfnamefont {C.}~\bibnamefont {Hofrichter}},
  \bibinfo {author} {\bibfnamefont {M.}~\bibnamefont {H{\"o}fer}}, \bibinfo
  {author} {\bibfnamefont {P.~C.}\ \bibnamefont {De~Groot}}, \bibinfo {author}
  {\bibfnamefont {I.}~\bibnamefont {Bloch}}, \ and\ \bibinfo {author}
  {\bibfnamefont {S.}~\bibnamefont {F{\"o}lling}},\ }\href {\doibase
  10.1038/nphys3061} {\bibfield  {journal} {\bibinfo  {journal} {Nature
  Physics}\ }\textbf {\bibinfo {volume} {10}},\ \bibinfo {pages} {779}
  (\bibinfo {year} {2014})}\BibitemShut {NoStop}%
\bibitem [{\citenamefont {Zhang}\ \emph {et~al.}(2014)\citenamefont {Zhang},
  \citenamefont {Bishof}, \citenamefont {Bromley}, \citenamefont {Kraus},
  \citenamefont {Safronova}, \citenamefont {Zoller}, \citenamefont {Rey},\ and\
  \citenamefont {Ye}}]{zhang_spectroscopic_2014}%
  \BibitemOpen
  \bibfield  {author} {\bibinfo {author} {\bibfnamefont {X.}~\bibnamefont
  {Zhang}}, \bibinfo {author} {\bibfnamefont {M.}~\bibnamefont {Bishof}},
  \bibinfo {author} {\bibfnamefont {S.~L.}\ \bibnamefont {Bromley}}, \bibinfo
  {author} {\bibfnamefont {C.~V.}\ \bibnamefont {Kraus}}, \bibinfo {author}
  {\bibfnamefont {M.~S.}\ \bibnamefont {Safronova}}, \bibinfo {author}
  {\bibfnamefont {P.}~\bibnamefont {Zoller}}, \bibinfo {author} {\bibfnamefont
  {A.~M.}\ \bibnamefont {Rey}}, \ and\ \bibinfo {author} {\bibfnamefont
  {J.}~\bibnamefont {Ye}},\ }\href {\doibase 10.1126/science.1254978}
  {\bibfield  {journal} {\bibinfo  {journal} {Science}\ }\textbf {\bibinfo
  {volume} {345}},\ \bibinfo {pages} {1467} (\bibinfo {year}
  {2014})}\BibitemShut {NoStop}%
\bibitem [{\citenamefont {Hofrichter}\ \emph {et~al.}(2016)\citenamefont
  {Hofrichter}, \citenamefont {Riegger}, \citenamefont {Scazza}, \citenamefont
  {H\"ofer}, \citenamefont {Fernandes}, \citenamefont {Bloch},\ and\
  \citenamefont {F\"olling}}]{hofrichter_direct_2016}%
  \BibitemOpen
  \bibfield  {author} {\bibinfo {author} {\bibfnamefont {C.}~\bibnamefont
  {Hofrichter}}, \bibinfo {author} {\bibfnamefont {L.}~\bibnamefont {Riegger}},
  \bibinfo {author} {\bibfnamefont {F.}~\bibnamefont {Scazza}}, \bibinfo
  {author} {\bibfnamefont {M.}~\bibnamefont {H\"ofer}}, \bibinfo {author}
  {\bibfnamefont {D.~R.}\ \bibnamefont {Fernandes}}, \bibinfo {author}
  {\bibfnamefont {I.}~\bibnamefont {Bloch}}, \ and\ \bibinfo {author}
  {\bibfnamefont {S.}~\bibnamefont {F\"olling}},\ }\href {\doibase
  10.1103/PhysRevX.6.021030} {\bibfield  {journal} {\bibinfo  {journal} {Phys.
  Rev. X}\ }\textbf {\bibinfo {volume} {6}},\ \bibinfo {pages} {021030}
  (\bibinfo {year} {2016})}\BibitemShut {NoStop}%
\bibitem [{\citenamefont {Assaad}(2005)}]{assaad_phase_2005}%
  \BibitemOpen
  \bibfield  {author} {\bibinfo {author} {\bibfnamefont {F.~F.}\ \bibnamefont
  {Assaad}},\ }\href {\doibase 10.1103/PhysRevB.71.075103} {\bibfield
  {journal} {\bibinfo  {journal} {Phys. Rev. B}\ }\textbf {\bibinfo {volume}
  {71}},\ \bibinfo {pages} {075103} (\bibinfo {year} {2005})}\BibitemShut
  {NoStop}%
\bibitem [{\citenamefont {Hermele}\ \emph {et~al.}(2009)\citenamefont
  {Hermele}, \citenamefont {Gurarie},\ and\ \citenamefont
  {Rey}}]{hermele_mott_2009}%
  \BibitemOpen
  \bibfield  {author} {\bibinfo {author} {\bibfnamefont {M.}~\bibnamefont
  {Hermele}}, \bibinfo {author} {\bibfnamefont {V.}~\bibnamefont {Gurarie}}, \
  and\ \bibinfo {author} {\bibfnamefont {A.~M.}\ \bibnamefont {Rey}},\ }\href
  {\doibase 10.1103/PhysRevLett.103.135301} {\bibfield  {journal} {\bibinfo
  {journal} {Phys. Rev. Lett.}\ }\textbf {\bibinfo {volume} {103}},\ \bibinfo
  {pages} {135301} (\bibinfo {year} {2009})}\BibitemShut {NoStop}%
\bibitem [{\citenamefont {Corboz}\ \emph {et~al.}(2011)\citenamefont {Corboz},
  \citenamefont {L\"auchli}, \citenamefont {Penc}, \citenamefont {Troyer},\
  and\ \citenamefont {Mila}}]{corboz_simultaneous_2011}%
  \BibitemOpen
  \bibfield  {author} {\bibinfo {author} {\bibfnamefont {P.}~\bibnamefont
  {Corboz}}, \bibinfo {author} {\bibfnamefont {A.~M.}\ \bibnamefont
  {L\"auchli}}, \bibinfo {author} {\bibfnamefont {K.}~\bibnamefont {Penc}},
  \bibinfo {author} {\bibfnamefont {M.}~\bibnamefont {Troyer}}, \ and\ \bibinfo
  {author} {\bibfnamefont {F.}~\bibnamefont {Mila}},\ }\href {\doibase
  10.1103/PhysRevLett.107.215301} {\bibfield  {journal} {\bibinfo  {journal}
  {Phys. Rev. Lett.}\ }\textbf {\bibinfo {volume} {107}},\ \bibinfo {pages}
  {215301} (\bibinfo {year} {2011})}\BibitemShut {NoStop}%
\bibitem [{\citenamefont {Hermele}\ and\ \citenamefont
  {Gurarie}(2011)}]{hermele_topological_2011}%
  \BibitemOpen
  \bibfield  {author} {\bibinfo {author} {\bibfnamefont {M.}~\bibnamefont
  {Hermele}}\ and\ \bibinfo {author} {\bibfnamefont {V.}~\bibnamefont
  {Gurarie}},\ }\href {\doibase 10.1103/PhysRevB.84.174441} {\bibfield
  {journal} {\bibinfo  {journal} {Phys. Rev. B}\ }\textbf {\bibinfo {volume}
  {84}},\ \bibinfo {pages} {174441} (\bibinfo {year} {2011})}\BibitemShut
  {NoStop}%
\bibitem [{\citenamefont {Corboz}\ \emph {et~al.}(2012)\citenamefont {Corboz},
  \citenamefont {Lajk\'o}, \citenamefont {L\"auchli}, \citenamefont {Penc},\
  and\ \citenamefont {Mila}}]{corboz_spin-orbital_2012}%
  \BibitemOpen
  \bibfield  {author} {\bibinfo {author} {\bibfnamefont {P.}~\bibnamefont
  {Corboz}}, \bibinfo {author} {\bibfnamefont {M.}~\bibnamefont {Lajk\'o}},
  \bibinfo {author} {\bibfnamefont {A.~M.}\ \bibnamefont {L\"auchli}}, \bibinfo
  {author} {\bibfnamefont {K.}~\bibnamefont {Penc}}, \ and\ \bibinfo {author}
  {\bibfnamefont {F.}~\bibnamefont {Mila}},\ }\href {\doibase
  10.1103/PhysRevX.2.041013} {\bibfield  {journal} {\bibinfo  {journal} {Phys.
  Rev. X}\ }\textbf {\bibinfo {volume} {2}},\ \bibinfo {pages} {041013}
  (\bibinfo {year} {2012})}\BibitemShut {NoStop}%
\bibitem [{\citenamefont {Capponi}\ \emph {et~al.}(2016)\citenamefont
  {Capponi}, \citenamefont {Lecheminant},\ and\ \citenamefont
  {Totsuka}}]{capponi_phases_2016}%
  \BibitemOpen
  \bibfield  {author} {\bibinfo {author} {\bibfnamefont {S.}~\bibnamefont
  {Capponi}}, \bibinfo {author} {\bibfnamefont {P.}~\bibnamefont
  {Lecheminant}}, \ and\ \bibinfo {author} {\bibfnamefont {K.}~\bibnamefont
  {Totsuka}},\ }\href {\doibase 10.1016/j.aop.2016.01.011} {\bibfield
  {journal} {\bibinfo  {journal} {Annals of Physics}\ }\textbf {\bibinfo
  {volume} {367}},\ \bibinfo {pages} {50} (\bibinfo {year} {2016})}\BibitemShut
  {NoStop}%
\bibitem [{\citenamefont {Dufour}\ and\ \citenamefont
  {Mila}(2016)}]{dufour_stabilization_2016}%
  \BibitemOpen
  \bibfield  {author} {\bibinfo {author} {\bibfnamefont {J.}~\bibnamefont
  {Dufour}}\ and\ \bibinfo {author} {\bibfnamefont {F.}~\bibnamefont {Mila}},\
  }\href {\doibase 10.1103/PhysRevA.94.033617} {\bibfield  {journal} {\bibinfo
  {journal} {Phys. Rev. A}\ }\textbf {\bibinfo {volume} {94}},\ \bibinfo
  {pages} {033617} (\bibinfo {year} {2016})}\BibitemShut {NoStop}%
\bibitem [{\citenamefont {Nataf}\ \emph
  {et~al.}(2016{\natexlab{a}})\citenamefont {Nataf}, \citenamefont {Lajk\'o},
  \citenamefont {Corboz}, \citenamefont {L\"auchli}, \citenamefont {Penc},\
  and\ \citenamefont {Mila}}]{nataf_plaquette_2016}%
  \BibitemOpen
  \bibfield  {author} {\bibinfo {author} {\bibfnamefont {P.}~\bibnamefont
  {Nataf}}, \bibinfo {author} {\bibfnamefont {M.}~\bibnamefont {Lajk\'o}},
  \bibinfo {author} {\bibfnamefont {P.}~\bibnamefont {Corboz}}, \bibinfo
  {author} {\bibfnamefont {A.~M.}\ \bibnamefont {L\"auchli}}, \bibinfo {author}
  {\bibfnamefont {K.}~\bibnamefont {Penc}}, \ and\ \bibinfo {author}
  {\bibfnamefont {F.}~\bibnamefont {Mila}},\ }\href {\doibase
  10.1103/PhysRevB.93.201113} {\bibfield  {journal} {\bibinfo  {journal} {Phys.
  Rev. B}\ }\textbf {\bibinfo {volume} {93}},\ \bibinfo {pages} {201113(R)}
  (\bibinfo {year} {2016}{\natexlab{a}})}\BibitemShut {NoStop}%
\bibitem [{\citenamefont {Nataf}\ \emph
  {et~al.}(2016{\natexlab{b}})\citenamefont {Nataf}, \citenamefont {Lajk\'o},
  \citenamefont {Wietek}, \citenamefont {Penc}, \citenamefont {Mila},\ and\
  \citenamefont {L\"auchli}}]{nataf_chiral_2016}%
  \BibitemOpen
  \bibfield  {author} {\bibinfo {author} {\bibfnamefont {P.}~\bibnamefont
  {Nataf}}, \bibinfo {author} {\bibfnamefont {M.}~\bibnamefont {Lajk\'o}},
  \bibinfo {author} {\bibfnamefont {A.}~\bibnamefont {Wietek}}, \bibinfo
  {author} {\bibfnamefont {K.}~\bibnamefont {Penc}}, \bibinfo {author}
  {\bibfnamefont {F.}~\bibnamefont {Mila}}, \ and\ \bibinfo {author}
  {\bibfnamefont {A.~M.}\ \bibnamefont {L\"auchli}},\ }\href {\doibase
  10.1103/PhysRevLett.117.167202} {\bibfield  {journal} {\bibinfo  {journal}
  {Phys. Rev. Lett.}\ }\textbf {\bibinfo {volume} {117}},\ \bibinfo {pages}
  {167202} (\bibinfo {year} {2016}{\natexlab{b}})}\BibitemShut {NoStop}%
\bibitem [{\citenamefont {Lecheminant}\ and\ \citenamefont
  {Tsvelik}(2017)}]{lecheminant_lattice_2017}%
  \BibitemOpen
  \bibfield  {author} {\bibinfo {author} {\bibfnamefont {P.}~\bibnamefont
  {Lecheminant}}\ and\ \bibinfo {author} {\bibfnamefont {A.~M.}\ \bibnamefont
  {Tsvelik}},\ }\href {\doibase 10.1103/PhysRevB.95.140406} {\bibfield
  {journal} {\bibinfo  {journal} {Phys. Rev. B}\ }\textbf {\bibinfo {volume}
  {95}},\ \bibinfo {pages} {140406(R)} (\bibinfo {year} {2017})}\BibitemShut
  {NoStop}%
\bibitem [{\citenamefont {Fuji}\ and\ \citenamefont
  {Lecheminant}(2017)}]{fuji_non-abelian_2017}%
  \BibitemOpen
  \bibfield  {author} {\bibinfo {author} {\bibfnamefont {Y.}~\bibnamefont
  {Fuji}}\ and\ \bibinfo {author} {\bibfnamefont {P.}~\bibnamefont
  {Lecheminant}},\ }\href {\doibase 10.1103/PhysRevB.95.125130} {\bibfield
  {journal} {\bibinfo  {journal} {Phys. Rev. B}\ }\textbf {\bibinfo {volume}
  {95}},\ \bibinfo {pages} {125130} (\bibinfo {year} {2017})}\BibitemShut
  {NoStop}%
\bibitem [{\citenamefont {Sutherland}(1975)}]{sutherland_model_1975}%
  \BibitemOpen
  \bibfield  {author} {\bibinfo {author} {\bibfnamefont {B.}~\bibnamefont
  {Sutherland}},\ }\href {\doibase 10.1103/PhysRevB.12.3795} {\bibfield
  {journal} {\bibinfo  {journal} {Phys. Rev. B}\ }\textbf {\bibinfo {volume}
  {12}},\ \bibinfo {pages} {3795} (\bibinfo {year} {1975})}\BibitemShut
  {NoStop}%
\bibitem [{\citenamefont {Affleck}\ and\ \citenamefont
  {Marston}(1988)}]{affleck_large-n_1988}%
  \BibitemOpen
  \bibfield  {author} {\bibinfo {author} {\bibfnamefont {I.}~\bibnamefont
  {Affleck}}\ and\ \bibinfo {author} {\bibfnamefont {J.~B.}\ \bibnamefont
  {Marston}},\ }\href {\doibase 10.1103/PhysRevB.37.3774} {\bibfield  {journal}
  {\bibinfo  {journal} {Phys. Rev. B}\ }\textbf {\bibinfo {volume} {37}},\
  \bibinfo {pages} {3774} (\bibinfo {year} {1988})}\BibitemShut {NoStop}%
\bibitem [{\citenamefont {Marston}\ and\ \citenamefont
  {Affleck}(1989)}]{marston_large-$n$_1989}%
  \BibitemOpen
  \bibfield  {author} {\bibinfo {author} {\bibfnamefont {J.~B.}\ \bibnamefont
  {Marston}}\ and\ \bibinfo {author} {\bibfnamefont {I.}~\bibnamefont
  {Affleck}},\ }\href {\doibase 10.1103/PhysRevB.39.11538} {\bibfield
  {journal} {\bibinfo  {journal} {Phys. Rev. B}\ }\textbf {\bibinfo {volume}
  {39}},\ \bibinfo {pages} {11538} (\bibinfo {year} {1989})}\BibitemShut
  {NoStop}%
\bibitem [{\citenamefont {Read}\ and\ \citenamefont
  {Sachdev}(1989{\natexlab{a}})}]{Read89}%
  \BibitemOpen
  \bibfield  {author} {\bibinfo {author} {\bibfnamefont {N.}~\bibnamefont
  {Read}}\ and\ \bibinfo {author} {\bibfnamefont {S.}~\bibnamefont {Sachdev}},\
  }\href {\doibase 10.1016/0550-3213(89)90061-8} {\bibfield  {journal}
  {\bibinfo  {journal} {Nuclear Physics B}\ }\textbf {\bibinfo {volume}
  {316}},\ \bibinfo {pages} {609 } (\bibinfo {year}
  {1989}{\natexlab{a}})}\BibitemShut {NoStop}%
\bibitem [{\citenamefont {Read}\ and\ \citenamefont
  {Sachdev}(1989{\natexlab{b}})}]{read_valence-bond_1989}%
  \BibitemOpen
  \bibfield  {author} {\bibinfo {author} {\bibfnamefont {N.}~\bibnamefont
  {Read}}\ and\ \bibinfo {author} {\bibfnamefont {S.}~\bibnamefont {Sachdev}},\
  }\href {\doibase 10.1103/PhysRevLett.62.1694} {\bibfield  {journal} {\bibinfo
   {journal} {Phys. Rev. Lett.}\ }\textbf {\bibinfo {volume} {62}},\ \bibinfo
  {pages} {1694} (\bibinfo {year} {1989}{\natexlab{b}})}\BibitemShut {NoStop}%
\bibitem [{\citenamefont {Read}\ and\ \citenamefont {Sachdev}(1990)}]{Read90}%
  \BibitemOpen
  \bibfield  {author} {\bibinfo {author} {\bibfnamefont {N.}~\bibnamefont
  {Read}}\ and\ \bibinfo {author} {\bibfnamefont {S.}~\bibnamefont {Sachdev}},\
  }\href {\doibase 10.1103/PhysRevB.42.4568} {\bibfield  {journal} {\bibinfo
  {journal} {Phys. Rev. B}\ }\textbf {\bibinfo {volume} {42}},\ \bibinfo
  {pages} {4568} (\bibinfo {year} {1990})}\BibitemShut {NoStop}%
\bibitem [{\citenamefont {Read}\ and\ \citenamefont
  {Sachdev}(1991)}]{read_large-_1991}%
  \BibitemOpen
  \bibfield  {author} {\bibinfo {author} {\bibfnamefont {N.}~\bibnamefont
  {Read}}\ and\ \bibinfo {author} {\bibfnamefont {S.}~\bibnamefont {Sachdev}},\
  }\href {\doibase 10.1103/PhysRevLett.66.1773} {\bibfield  {journal} {\bibinfo
   {journal} {Phys. Rev. Lett.}\ }\textbf {\bibinfo {volume} {66}},\ \bibinfo
  {pages} {1773} (\bibinfo {year} {1991})}\BibitemShut {NoStop}%
\bibitem [{\citenamefont {Onufriev}\ and\ \citenamefont
  {Marston}(1999)}]{onufriev_enlarged_1999}%
  \BibitemOpen
  \bibfield  {author} {\bibinfo {author} {\bibfnamefont {A.~V.}\ \bibnamefont
  {Onufriev}}\ and\ \bibinfo {author} {\bibfnamefont {J.~B.}\ \bibnamefont
  {Marston}},\ }\href {\doibase 10.1103/PhysRevB.59.12573} {\bibfield
  {journal} {\bibinfo  {journal} {Phys. Rev. B}\ }\textbf {\bibinfo {volume}
  {59}},\ \bibinfo {pages} {12573} (\bibinfo {year} {1999})}\BibitemShut
  {NoStop}%
\bibitem [{\citenamefont {Assaraf}\ \emph {et~al.}(2004)\citenamefont
  {Assaraf}, \citenamefont {Azaria}, \citenamefont {Boulat}, \citenamefont
  {Caffarel},\ and\ \citenamefont {Lecheminant}}]{assaraf_dynamical_2004}%
  \BibitemOpen
  \bibfield  {author} {\bibinfo {author} {\bibfnamefont {R.}~\bibnamefont
  {Assaraf}}, \bibinfo {author} {\bibfnamefont {P.}~\bibnamefont {Azaria}},
  \bibinfo {author} {\bibfnamefont {E.}~\bibnamefont {Boulat}}, \bibinfo
  {author} {\bibfnamefont {M.}~\bibnamefont {Caffarel}}, \ and\ \bibinfo
  {author} {\bibfnamefont {P.}~\bibnamefont {Lecheminant}},\ }\href {\doibase
  10.1103/PhysRevLett.93.016407} {\bibfield  {journal} {\bibinfo  {journal}
  {Phys. Rev. Lett.}\ }\textbf {\bibinfo {volume} {93}},\ \bibinfo {pages}
  {016407} (\bibinfo {year} {2004})}\BibitemShut {NoStop}%
\bibitem [{\citenamefont {Paramekanti}\ and\ \citenamefont
  {Marston}(2007)}]{paramekanti_su_2007}%
  \BibitemOpen
  \bibfield  {author} {\bibinfo {author} {\bibfnamefont {A.}~\bibnamefont
  {Paramekanti}}\ and\ \bibinfo {author} {\bibfnamefont {J.~B.}\ \bibnamefont
  {Marston}},\ }\href {\doibase 10.1088/0953-8984/19/12/125215} {\bibfield
  {journal} {\bibinfo  {journal} {Journal of Physics: Condensed Matter}\
  }\textbf {\bibinfo {volume} {19}},\ \bibinfo {pages} {125215} (\bibinfo
  {year} {2007})}\BibitemShut {NoStop}%
\bibitem [{\citenamefont {Nonne}\ \emph {et~al.}(2011)\citenamefont {Nonne},
  \citenamefont {Lecheminant}, \citenamefont {Capponi}, \citenamefont {Roux},\
  and\ \citenamefont {Boulat}}]{nonne_competing_2011}%
  \BibitemOpen
  \bibfield  {author} {\bibinfo {author} {\bibfnamefont {H.}~\bibnamefont
  {Nonne}}, \bibinfo {author} {\bibfnamefont {P.}~\bibnamefont {Lecheminant}},
  \bibinfo {author} {\bibfnamefont {S.}~\bibnamefont {Capponi}}, \bibinfo
  {author} {\bibfnamefont {G.}~\bibnamefont {Roux}}, \ and\ \bibinfo {author}
  {\bibfnamefont {E.}~\bibnamefont {Boulat}},\ }\href {\doibase
  10.1103/PhysRevB.84.125123} {\bibfield  {journal} {\bibinfo  {journal} {Phys.
  Rev. B}\ }\textbf {\bibinfo {volume} {84}},\ \bibinfo {pages} {125123}
  (\bibinfo {year} {2011})}\BibitemShut {NoStop}%
\bibitem [{\citenamefont {Mermin}\ and\ \citenamefont
  {Wagner}(1966)}]{mermin_absence_1966}%
  \BibitemOpen
  \bibfield  {author} {\bibinfo {author} {\bibfnamefont {N.~D.}\ \bibnamefont
  {Mermin}}\ and\ \bibinfo {author} {\bibfnamefont {H.}~\bibnamefont
  {Wagner}},\ }\href {\doibase 10.1103/PhysRevLett.17.1133} {\bibfield
  {journal} {\bibinfo  {journal} {Phys. Rev. Lett.}\ }\textbf {\bibinfo
  {volume} {17}},\ \bibinfo {pages} {1133} (\bibinfo {year}
  {1966})}\BibitemShut {NoStop}%
\bibitem [{\citenamefont {Coleman}(1973)}]{coleman_there_1973}%
  \BibitemOpen
  \bibfield  {author} {\bibinfo {author} {\bibfnamefont {S.}~\bibnamefont
  {Coleman}},\ }\href {\doibase 10.1007/BF01646487} {\bibfield  {journal}
  {\bibinfo  {journal} {Communications in Mathematical Physics}\ }\textbf
  {\bibinfo {volume} {31}},\ \bibinfo {pages} {259} (\bibinfo {year}
  {1973})}\BibitemShut {NoStop}%
\bibitem [{\citenamefont {Kim}\ \emph {et~al.}(2017)\citenamefont {Kim},
  \citenamefont {Penc}, \citenamefont {Nataf},\ and\ \citenamefont
  {Mila}}]{kim_linear_2017}%
  \BibitemOpen
  \bibfield  {author} {\bibinfo {author} {\bibfnamefont {F.~H.}\ \bibnamefont
  {Kim}}, \bibinfo {author} {\bibfnamefont {K.}~\bibnamefont {Penc}}, \bibinfo
  {author} {\bibfnamefont {P.}~\bibnamefont {Nataf}}, \ and\ \bibinfo {author}
  {\bibfnamefont {F.}~\bibnamefont {Mila}},\ }\href {\doibase
  10.1103/PhysRevB.96.205142} {\bibfield  {journal} {\bibinfo  {journal} {Phys.
  Rev. B}\ }\textbf {\bibinfo {volume} {96}},\ \bibinfo {pages} {205142}
  (\bibinfo {year} {2017})}\BibitemShut {NoStop}%
\bibitem [{\citenamefont {Cai}\ \emph {et~al.}(2013)\citenamefont {Cai},
  \citenamefont {Hung}, \citenamefont {Wang},\ and\ \citenamefont
  {Wu}}]{cai_quantum_2013}%
  \BibitemOpen
  \bibfield  {author} {\bibinfo {author} {\bibfnamefont {Z.}~\bibnamefont
  {Cai}}, \bibinfo {author} {\bibfnamefont {H.-H.}\ \bibnamefont {Hung}},
  \bibinfo {author} {\bibfnamefont {L.}~\bibnamefont {Wang}}, \ and\ \bibinfo
  {author} {\bibfnamefont {C.}~\bibnamefont {Wu}},\ }\href {\doibase
  10.1103/PhysRevB.88.125108} {\bibfield  {journal} {\bibinfo  {journal} {Phys.
  Rev. B}\ }\textbf {\bibinfo {volume} {88}},\ \bibinfo {pages} {125108}
  (\bibinfo {year} {2013})}\BibitemShut {NoStop}%
\bibitem [{\citenamefont {Wang}\ \emph {et~al.}(2014)\citenamefont {Wang},
  \citenamefont {Li}, \citenamefont {Cai}, \citenamefont {Zhou}, \citenamefont
  {Wang},\ and\ \citenamefont {Wu}}]{Wang14b}%
  \BibitemOpen
  \bibfield  {author} {\bibinfo {author} {\bibfnamefont {D.}~\bibnamefont
  {Wang}}, \bibinfo {author} {\bibfnamefont {Y.}~\bibnamefont {Li}}, \bibinfo
  {author} {\bibfnamefont {Z.}~\bibnamefont {Cai}}, \bibinfo {author}
  {\bibfnamefont {Z.}~\bibnamefont {Zhou}}, \bibinfo {author} {\bibfnamefont
  {Y.}~\bibnamefont {Wang}}, \ and\ \bibinfo {author} {\bibfnamefont
  {C.}~\bibnamefont {Wu}},\ }\href {\doibase 10.1103/PhysRevLett.112.156403}
  {\bibfield  {journal} {\bibinfo  {journal} {Phys. Rev. Lett.}\ }\textbf
  {\bibinfo {volume} {112}},\ \bibinfo {pages} {156403} (\bibinfo {year}
  {2014})}\BibitemShut {NoStop}%
\bibitem [{\citenamefont {Matsumoto}\ \emph {et~al.}(2010)\citenamefont
  {Matsumoto}, \citenamefont {Kuroe}, \citenamefont {Sekine},\ and\
  \citenamefont {Masuda}}]{masashige_transverse_2010}%
  \BibitemOpen
  \bibfield  {author} {\bibinfo {author} {\bibfnamefont {M.}~\bibnamefont
  {Matsumoto}}, \bibinfo {author} {\bibfnamefont {H.}~\bibnamefont {Kuroe}},
  \bibinfo {author} {\bibfnamefont {T.}~\bibnamefont {Sekine}}, \ and\ \bibinfo
  {author} {\bibfnamefont {T.}~\bibnamefont {Masuda}},\ }\href {\doibase
  10.1143/JPSJ.79.084703} {\bibfield  {journal} {\bibinfo  {journal} {Journal
  of the Physical Society of Japan}\ }\textbf {\bibinfo {volume} {79}},\
  \bibinfo {pages} {084703} (\bibinfo {year} {2010})}\BibitemShut {NoStop}%
\bibitem [{\citenamefont {Romh\'anyi}\ and\ \citenamefont
  {Penc}(2012)}]{romhanyi_multiboson_2012}%
  \BibitemOpen
  \bibfield  {author} {\bibinfo {author} {\bibfnamefont {J.}~\bibnamefont
  {Romh\'anyi}}\ and\ \bibinfo {author} {\bibfnamefont {K.}~\bibnamefont
  {Penc}},\ }\href {\doibase 10.1103/PhysRevB.86.174428} {\bibfield  {journal}
  {\bibinfo  {journal} {Phys. Rev. B}\ }\textbf {\bibinfo {volume} {86}},\
  \bibinfo {pages} {174428} (\bibinfo {year} {2012})}\BibitemShut {NoStop}%
\bibitem [{\citenamefont {Penc}\ \emph {et~al.}(2012)\citenamefont {Penc},
  \citenamefont {Romh\'anyi}, \citenamefont {R{\~o}{\~o}m}, \citenamefont
  {Nagel}, \citenamefont {Antal}, \citenamefont {Feh\'er}, \citenamefont
  {J\'anossy}, \citenamefont {Engelkamp}, \citenamefont {Murakawa},
  \citenamefont {Tokura}, \citenamefont {Szaller}, \citenamefont {Bord\'acs},\
  and\ \citenamefont {K\'ezsm\'arki}}]{penc_spin-stretching_2012}%
  \BibitemOpen
  \bibfield  {author} {\bibinfo {author} {\bibfnamefont {K.}~\bibnamefont
  {Penc}}, \bibinfo {author} {\bibfnamefont {J.}~\bibnamefont {Romh\'anyi}},
  \bibinfo {author} {\bibfnamefont {T.}~\bibnamefont {R{\~o}{\~o}m}}, \bibinfo
  {author} {\bibfnamefont {U.}~\bibnamefont {Nagel}}, \bibinfo {author}
  {\bibfnamefont {A.}~\bibnamefont {Antal}}, \bibinfo {author} {\bibfnamefont
  {T.}~\bibnamefont {Feh\'er}}, \bibinfo {author} {\bibfnamefont
  {A.}~\bibnamefont {J\'anossy}}, \bibinfo {author} {\bibfnamefont
  {H.}~\bibnamefont {Engelkamp}}, \bibinfo {author} {\bibfnamefont
  {H.}~\bibnamefont {Murakawa}}, \bibinfo {author} {\bibfnamefont
  {Y.}~\bibnamefont {Tokura}}, \bibinfo {author} {\bibfnamefont
  {D.}~\bibnamefont {Szaller}}, \bibinfo {author} {\bibfnamefont
  {S.}~\bibnamefont {Bord\'acs}}, \ and\ \bibinfo {author} {\bibfnamefont
  {I.}~\bibnamefont {K\'ezsm\'arki}},\ }\href {\doibase
  10.1103/PhysRevLett.108.257203} {\bibfield  {journal} {\bibinfo  {journal}
  {Phys. Rev. Lett.}\ }\textbf {\bibinfo {volume} {108}},\ \bibinfo {pages}
  {257203} (\bibinfo {year} {2012})}\BibitemShut {NoStop}%
\bibitem [{\citenamefont {Bercx}\ \emph {et~al.}(2017)\citenamefont {Bercx},
  \citenamefont {Goth}, \citenamefont {Hofmann},\ and\ \citenamefont
  {Assaad}}]{ALF_v1}%
  \BibitemOpen
  \bibfield  {author} {\bibinfo {author} {\bibfnamefont {M.}~\bibnamefont
  {Bercx}}, \bibinfo {author} {\bibfnamefont {F.}~\bibnamefont {Goth}},
  \bibinfo {author} {\bibfnamefont {J.~S.}\ \bibnamefont {Hofmann}}, \ and\
  \bibinfo {author} {\bibfnamefont {F.~F.}\ \bibnamefont {Assaad}},\ }\href
  {\doibase 10.21468/SciPostPhys.3.2.013} {\bibfield  {journal} {\bibinfo
  {journal} {SciPost Phys.}\ }\textbf {\bibinfo {volume} {3}},\ \bibinfo
  {pages} {013} (\bibinfo {year} {2017})}\BibitemShut {NoStop}%
\bibitem [{\citenamefont {Haldane}(1988)}]{Haldane88}%
  \BibitemOpen
  \bibfield  {author} {\bibinfo {author} {\bibfnamefont {F.~D.~M.}\
  \bibnamefont {Haldane}},\ }\href {\doibase 10.1103/PhysRevLett.61.1029}
  {\bibfield  {journal} {\bibinfo  {journal} {Phys. Rev. Lett.}\ }\textbf
  {\bibinfo {volume} {61}},\ \bibinfo {pages} {1029} (\bibinfo {year}
  {1988})}\BibitemShut {NoStop}%
\bibitem [{\citenamefont {Senthil}\ \emph
  {et~al.}(2004{\natexlab{a}})\citenamefont {Senthil}, \citenamefont
  {Vishwanath}, \citenamefont {Balents}, \citenamefont {Sachdev},\ and\
  \citenamefont {Fisher}}]{Senthil1490}%
  \BibitemOpen
  \bibfield  {author} {\bibinfo {author} {\bibfnamefont {T.}~\bibnamefont
  {Senthil}}, \bibinfo {author} {\bibfnamefont {A.}~\bibnamefont {Vishwanath}},
  \bibinfo {author} {\bibfnamefont {L.}~\bibnamefont {Balents}}, \bibinfo
  {author} {\bibfnamefont {S.}~\bibnamefont {Sachdev}}, \ and\ \bibinfo
  {author} {\bibfnamefont {M.~P.~A.}\ \bibnamefont {Fisher}},\ }\href {\doibase
  10.1126/science.1091806} {\bibfield  {journal} {\bibinfo  {journal}
  {Science}\ }\textbf {\bibinfo {volume} {303}},\ \bibinfo {pages} {1490}
  (\bibinfo {year} {2004}{\natexlab{a}})}\BibitemShut {NoStop}%
\bibitem [{\citenamefont {Senthil}\ \emph
  {et~al.}(2004{\natexlab{b}})\citenamefont {Senthil}, \citenamefont {Balents},
  \citenamefont {Sachdev}, \citenamefont {Vishwanath},\ and\ \citenamefont
  {Fisher}}]{Senthil04_1}%
  \BibitemOpen
  \bibfield  {author} {\bibinfo {author} {\bibfnamefont {T.}~\bibnamefont
  {Senthil}}, \bibinfo {author} {\bibfnamefont {L.}~\bibnamefont {Balents}},
  \bibinfo {author} {\bibfnamefont {S.}~\bibnamefont {Sachdev}}, \bibinfo
  {author} {\bibfnamefont {A.}~\bibnamefont {Vishwanath}}, \ and\ \bibinfo
  {author} {\bibfnamefont {M.~P.~A.}\ \bibnamefont {Fisher}},\ }\href {\doibase
  10.1103/PhysRevB.70.144407} {\bibfield  {journal} {\bibinfo  {journal} {Phys.
  Rev. B}\ }\textbf {\bibinfo {volume} {70}},\ \bibinfo {pages} {144407}
  (\bibinfo {year} {2004}{\natexlab{b}})}\BibitemShut {NoStop}%
\bibitem [{\citenamefont {Xu}\ \emph {et~al.}(2019)\citenamefont {Xu},
  \citenamefont {Qi}, \citenamefont {Zhang}, \citenamefont {Assaad},
  \citenamefont {Xu},\ and\ \citenamefont {Meng}}]{Xu18}%
  \BibitemOpen
  \bibfield  {author} {\bibinfo {author} {\bibfnamefont {X.~Y.}\ \bibnamefont
  {Xu}}, \bibinfo {author} {\bibfnamefont {Y.}~\bibnamefont {Qi}}, \bibinfo
  {author} {\bibfnamefont {L.}~\bibnamefont {Zhang}}, \bibinfo {author}
  {\bibfnamefont {F.~F.}\ \bibnamefont {Assaad}}, \bibinfo {author}
  {\bibfnamefont {C.}~\bibnamefont {Xu}}, \ and\ \bibinfo {author}
  {\bibfnamefont {Z.~Y.}\ \bibnamefont {Meng}},\ }\href {\doibase
  10.1103/PhysRevX.9.021022} {\bibfield  {journal} {\bibinfo  {journal} {Phys.
  Rev. X}\ }\textbf {\bibinfo {volume} {9}},\ \bibinfo {pages} {021022}
  (\bibinfo {year} {2019})}\BibitemShut {NoStop}%
\bibitem [{\citenamefont {Assaad}\ and\ \citenamefont
  {Herbut}(2013)}]{Assaad13}%
  \BibitemOpen
  \bibfield  {author} {\bibinfo {author} {\bibfnamefont {F.~F.}\ \bibnamefont
  {Assaad}}\ and\ \bibinfo {author} {\bibfnamefont {I.~F.}\ \bibnamefont
  {Herbut}},\ }\href {\doibase 10.1103/PhysRevX.3.031010} {\bibfield  {journal}
  {\bibinfo  {journal} {Phys. Rev. X}\ }\textbf {\bibinfo {volume} {3}},\
  \bibinfo {pages} {031010} (\bibinfo {year} {2013})}\BibitemShut {NoStop}%
\end{thebibliography}%

\end{document}